\documentclass[authoryear, preprint, 12pt, 3p]{elsarticle}

\journal{Neurocomputing}

\usepackage[T1]{fontenc} 
\usepackage{float}
\usepackage{amssymb, amsmath, amsthm}
\usepackage{caption}
\usepackage{hyperref}
\usepackage[table]{xcolor}
\usepackage{booktabs, multirow}
\usepackage[separate-uncertainty=true, %
            multi-part-units=single, %
            range-units=single]{siunitx}

\usepackage{lineno}

\begin{document}

\begin{frontmatter}



\title{Emergent complexity and rhythms in evoked and spontaneous dynamics of human whole-brain models after tuning through analysis tools} 

\author[unimi1,unimi2]{Gianluca Gaglioti\corref{cor1}\fnref{fn1}} 
\author[ucbm,infn]{Alessandra Cardinale\fnref{fn1}} 
\author[infn]{Cosimo Lupo\fnref{fn1}\corref{cor2}} 
\author[unimi3]{Thierry Nieus} 
\author[infn]{Federico Marmoreo} 
\author[unimi1,unimi2]{Elena Focacci} 
\author[nyu]{Robin Gutzen} 
\author[fzj]{Michael Denker} 
\author[unimi4,policlinicomi]{Andrea Pigorini} 
\author[unimi1,gnocchi,azrieli]{Marcello Massimini\fnref{fn2}} 
\author[unimi1]{Simone Sarasso\fnref{fn2}} 
\author[infn]{Pier Stanislao Paolucci\fnref{fn2}} 
\author[infn]{Giulia De Bonis\fnref{fn2}} 

\cortext[cor1]{Corresponding author. Email address: gaglioti.gianluca@gmail.com.}
\cortext[cor2]{Corresponding author. Email address: cosimo.lupo89@gmail.com.}
\fntext[fn1]{These authors contributed equally.}
\fntext[fn2]{These authors jointly supervised this work.}

\affiliation[unimi1]{organization={Dep. of Biomedical and Clinical Sciences, Univ. of Milan},
            city={Milan},
            country={Italy}}
\affiliation[unimi2]{organization={Dep. of Philosophy ``Piero Martinetti'', Univ. of Milan},
            city={Milan},
            country={Italy}}
\affiliation[ucbm]{organization={Universit\`a Campus Bio-Medico di Roma},
            city={Rome},
            country={Italy}}
\affiliation[infn]{organization={Istituto Nazionale di Fisica Nucleare (INFN), Sezione di Roma},
            city={Rome},
            country={Italy}}
\affiliation[unimi3]{organization={Dep. of Environmental Science and Policy, Univ. of Milan},
            city={Milan},
            country={Italy}}
\affiliation[nyu]{organization={Center for Data Science, New York University},
            city={New York},
            country={U.S.A.}}
\affiliation[fzj]{organization={Institute for Advanced Simulation (IAS-6), J\"ulich Research Centre},
            city={J\"ulich},
            country={Germany}}
\affiliation[unimi4]{organization={Dep. of Biomedical, Surgical and Dental Sciences, Univ. of Milan},
            city={Milan},
            country={Italy}}
\affiliation[policlinicomi]{organization={UOC Maxillo-facial Surgery and dentistry, Fondazione IRCCS C\`a Granda, Ospedale Maggiore Policlinico},
            city={Milan},
            country={Italy}}
\affiliation[gnocchi]{organization={IRCCS Fondazione Don Carlo Gnocchi ONLUS},
            city={Milan},
            country={Italy}}
\affiliation[azrieli]{organization={Azrieli Program in Brain, Mind and Consciousness, Canadian Institute for Advanced Research},
            city={Toronto},
            country={Canada}}

\begin{abstract}
    The simulation of whole-brain dynamics should reproduce realistic spontaneous and evoked neural activity across different scales, including emergent rhythms, spatio-temporal activation patterns, and macroscale complexity. Once a mathematical model is selected, its configuration must be determined by properly setting its parameters. A critical preliminary step in this process is defining an appropriate set of observables to guide the selection of model configurations (parameter tuning), laying the groundwork for quantitative calibration of accurate whole-brain models. Here, we address this challenge by presenting a framework that integrates two complementary tools: The Virtual Brain (TVB) platform for simulating whole-brain dynamics, and the Collaborative Brain Wave Analysis Pipeline (Cobrawap) for analyzing simulation outputs using a set of standardized metrics. We apply this framework to a 998-node human connectome, using two configurations of the Larter-Breakspear neural mass model: one with the TVB default parameters, the other tuned using Cobrawap. The results reveal that the tuned configuration exhibits several biologically relevant features, absent in the default model for both spontaneous and evoked dynamics. In response to external perturbations, the tuned model generates non-stereotyped, complex spatio-temporal activity, as measured by the perturbational complexity index. In spontaneous activity, it exhibits robust alpha-band oscillations, infra-slow rhythms, scale-free characteristics, greater spatio-temporal heterogeneity, and asymmetric functional connectivity. This work demonstrates how combining TVB and Cobrawap can guide parameter tuning and lays the groundwork for data-driven calibration and validation of accurate whole-brain models.
\end{abstract}





\end{frontmatter}


\section{Introduction}
\label{sec:intro}

Whole-brain computational models based on the neural mass approximation provide a framework for investigating the fundamental mechanisms underlying large-scale neural dynamics \citep{breakspear2017dynamic, griffiths2022wholebrain}. However, a crucial challenge remains: how to configure and adjust such a model so that it simultaneously captures key features of spontaneous and evoked brain activity, and ultimately match empirical data.
Among the essential features that a biologically realistic whole-brain model should reproduce, the alpha rhythm (\qtyrange[range-phrase=--]{8}{12}{\Hz}) stands out as a fundamental oscillation in large-scale brain dynamics. First described by \citet{berger1929uber}, it is closely associated with states of relaxed wakefulness and is thought to play a role in cognitive processes \citep{klimesch1999eeg, jensen2010shaping}. While alpha oscillations are a hallmark of resting-state dynamics, their fluctuating and multistable nature reflects the inherent variability and the transient state shifts of the brain \citep{freyer2009bistability, freyer2011biophysical}. In addition, they do not act in isolation but rather coexist with fluctuations occurring across multiple timescales \citep{fox2007spontaneous, palva2012infraslow}. At rest, brain activity fluctuates over time, exhibiting complex patterns of spontaneous variation in neuronal signals shaping the functional organization of the brain \citep{deco2017dynamics}. Fluctuations also exhibit significant regional heterogeneity, with varying degrees of signal variability influenced by anatomical connectivity and functional roles \citep{deco2011emerging}. High-fluctuation regions act as integrative hubs, crucial for coordinating information flow across the network \citep{rabuffo2021neuronal, vandenheuvel2013network}. These regions may contribute to infra-slow fluctuations ($ < \qty{0.1}{\Hz}$), which shape large-scale brain dynamics \citep{palva2012infraslow, gutierrezbarragan2019infraslow, rabuffo2021neuronal, vayrynen2023infraslow} and drive dynamic functional connectivity and brain state reconfiguration \citep{tagliazucchi2012dynamic, gutierrezbarragan2019infraslow}. Beyond this spontaneous dynamics, the awake brain also exhibits sustained and spatially complex responses evoked by external focal perturbations, a phenomenon that has been extensively studied in experimental settings, such as in transcranial magnetic stimulation combined with simultaneous electroencephalography (TMS-EEG) experiments, where the spatio-temporal complexity of the evoked cortical response has been shown to correlate with the level of consciousness \citep{rosanova2012recovery, casali2013theoretically, casarotto2016stratification}. 
In this context, various forms of external brain stimulation, including non-invasive neuromodulatory techniques, are capable of directly modulating these spontaneous and evoked features, producing measurable changes in both the periodic and aperiodic components of neural activity, as well as in functional connectivity patterns \citep{yu2018modulation, yu2019modulation, yu2024evaluation, liu2022acupuncture, mueller2023hdtdcs, gao2025transcranial, kar2020transcranial, masina2025transcranial}.

These features -- multistability, scale-free and infra-slow fluctuations, dynamic reconfiguration, and complex evoked responses -- have been increasingly linked to critical dynamics in the brain \citep{deco2012ongoing, obyrne2022how}. The hypothesis of brain criticality posits that neural systems operate near a critical point between order and disorder, a regime that supports maximal variability, long-range correlations, and optimal responsiveness to external inputs \citep{beggs2003neuronal, chialvo2010emergent, shew2012functional, cocchi2017criticality, palva2018roles, maschke2024critical}.

As anticipated, a key question emerging from these considerations is whether a single, unified model can simultaneously reproduce both features of spontaneous and evoked brain activity. Indeed, traditionally, these two have been treated as separate problems in computational modeling. However, a relationship is expected between the richness of spontaneous activity patterns and the ability of the brain to sustain complex responses to external perturbations through its network dynamics. Therefore, capturing the interplay of spontaneous and evoked dynamics, while assessing their consistency with empirical data, should be considered as a non-trivial requirement for the simulated network.

This assessment process grounds on an initial parameter \textit{tuning}, aimed at addressing the correct sub-region in the parameter space, exploring how changes in model settings influence target observables. The set of quantitative observables -- defined in the tuning to assess the quality of the model -- will be named from now on as \textit{metrics}. Tuning is followed by a \textit{calibration}, that adjusts the model parameters -- ideally using an automated optimization process -- by leveraging the metrics defined during the tuning to match a specific/personalized dataset. When discrepancies occur, these metrics offer feedback for refining the model and identifying its region of applicability, ensuring that the model reproduces not only isolated features, but also the brain broader dynamical landscape. It is worth noting that the identification of the starting parameter domain performed during the tuning is an essential ingredient for any automated calibration process that explores a high dimensional parameter space. Finally, a more thorough \textit{validation} should be performed against a range of statistical features to establish their agreement with respect to independent experimental data \citep[e.g.,][]{trensch2018rigorous, gutzen2018reproducible}, leveraging the wealth of heterogeneous data available and facilitating cross-study comparisons.
%
This tuning/calibration/validation procedure should rely on a processing of simulation outputs methodologically consistent with the analysis of experimental outcomes, especially when aiming at embracing a wider range of case studies for a more precise alignment between theoretical predictions and experimental observations.
In addition, to enhance the reproducibility and the scope of application of the simulations, the model setup and the metrics used for data analysis should be accessible to the broader neuroscientific community, standardized across the calibration/validation loops and carried out by a robust simulation/analysis workflow.

A number of frameworks address these needs for standardization, both for simulation and data analytics. Platforms like The Virtual Brain (TVB) \citep{sanzleon2013virtual, sanzleon2015mathematical}, the Brain Dynamics Toolbox \citep{heitmann2018brain}, and Neurolib \citep{cakan2023neurolib} facilitate and standardize whole-brain simulations using neural mass models, which describe large-scale neural interactions via phenomenological or mean-field approaches \citep{breakspear2017dynamic}. Among the advantages of these solutions, the prebuilt, validated models offered by these platforms streamline calibration against empirical data, allowing researchers to focus on scientific questions rather than custom coding. Active user communities further enhance reproducibility and collaboration.
Tools like MNE \citep{gramfort2013meg} for EEG/MEG, FreeSurfer \citep{fischl2012freesurfer} for Magnetic Resonance Imaging (MRI), fMRIPrep \citep{esteban2019fmriprep} and CPAC \citep{mainas2024exploring} for functional MRI (fMRI), and Elephant~\citep{denker2018collaborative} for electrophysiological data already provide reproducible analyses. To different extents, such methodologies integrate heterogeneous data, minimize inconsistencies, and enhance comparability, crucial for addressing the challenges of multimodal neuroscience. However, tools like the above are usually designed to address specific experimental cases, resulting in highly tailored solutions but also potentially reduced generalizability, which may require separate approaches when performing data versus model comparisons.
Extending this principle to complete analytics workflows, the Collaborative Brain Wave Analysis Pipeline (Cobrawap) \citep{gutzen2024modular, cobrawap_docs} is developed as an open-source tool providing standardized and quantitative descriptions of brain wave phenomena in both experimental recordings and in silico data \citep{capone2023simulations}, addressing the growing need for shared and agreed metrics and methodologies in the field. 
Complementing this ecosystem of tools, broad initiatives like the Human Connectome Project \citep{vanessen2012human}, the Allen Brain Atlas \citep{sunkin2013allen}, and EBRAINS \citep{ebrains_webpage} provide publicly available curated data and frameworks.

\begin{figure}[!b]
    \centering
    \includegraphics[width=\linewidth]{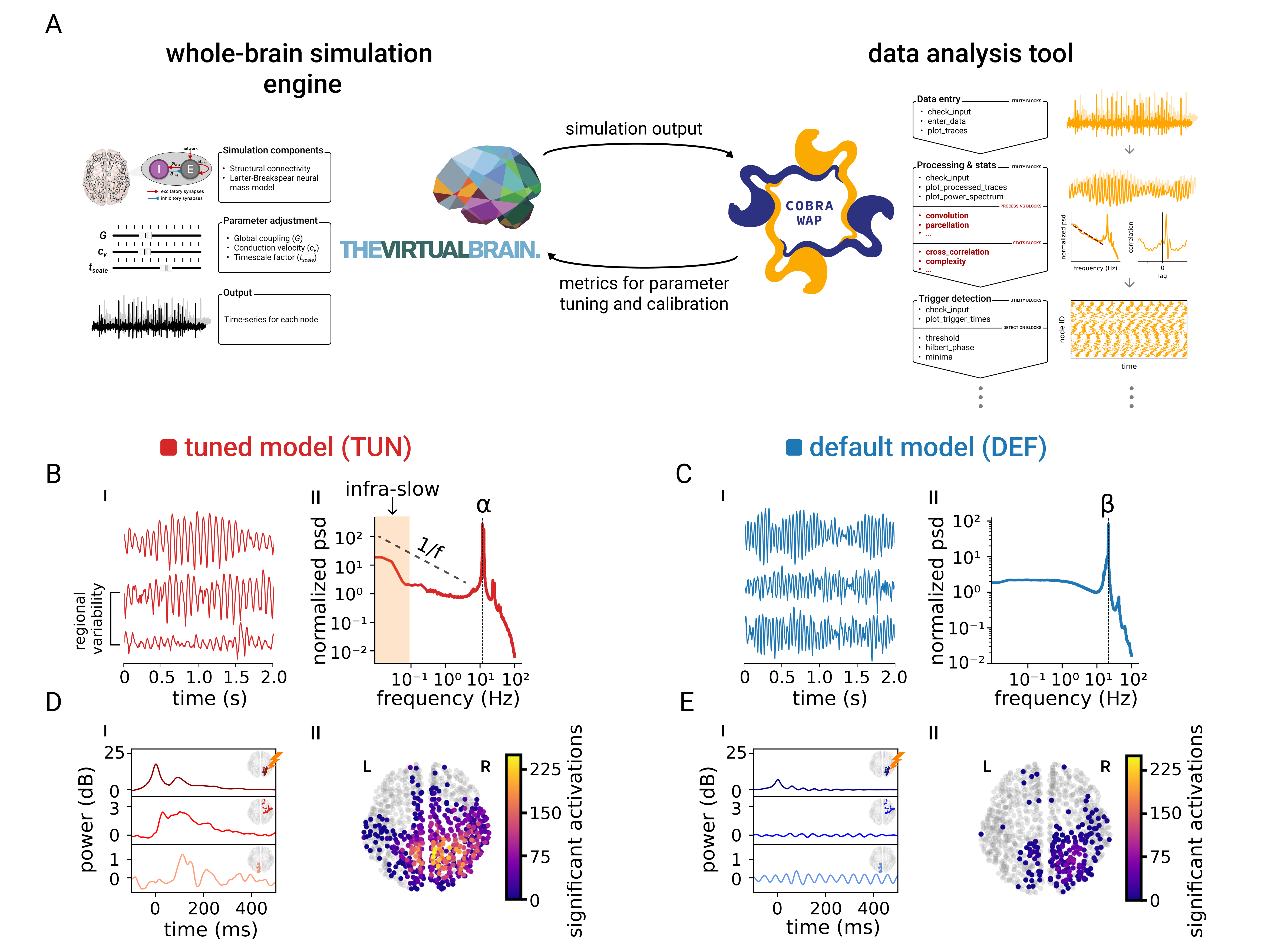}
    \caption{
        \textbf{TVB simulations and analysis with Cobrawap.}\\
        \textbf{A)} The TVB-Cobrawap workflow (center) facilitates tuning, calibration and validation of TVB model parameters; left and right diagrams sketch the sequence of actions requested, respectively, for launching TVB simulations and for processing TVB output with Cobrawap. The latter is adapted from \citet{gutzen2024modular} and illustrates the sequence of stages and blocks arranged along the pipeline for the study of brain dynamics; text in red highlights methods and metrics specifically implemented for addressing the TVB-tuning use-case, like cross correlations, parcellation reduction, complexity, in addition to what already provided by the Cobrawap pipeline -- such as trace statistics, spectral analysis, event detection. Plots are included in the diagrams as exemplary for illustration purposes; detailed discussion is given later in the manuscript.
     }
    \label{fig:fig_1}
\end{figure}

\begin{figure}[!t]
    \centering
    \ContinuedFloat
    \caption*{
        \textbf{B)} Illustrative examples of three spontaneous activity traces from three cortical regions of the tuned (TUN) configuration (B$_{\mathrm{I}}$). These represent proxies of synaptic activity, obtained by convolving the voltage traces with a bi-exponential kernel (see Sec.~\ref{subsubsec:mat_meth:metrics:PSD}) at the single-node level, and subsequently averaging the resulting signals across nodes belonging to each region. From top to bottom, the traces correspond to the average synaptic activity across left-hemisphere nodes belonging to the IP, IT, and PARC regions, respectively (see Suppl. Tab.~\ref{tab:regions}). The average Power Spectral Density (PSD) across all regions displays a clear peak in the alpha band, at approximately \qty{12}{\Hz}, together with an enriched infra-slow spectral component (B$_{\mathrm{II}}$).
        \textbf{C)} Same as B), referred to the default (DEF) configuration. In this case, the dominant peak in the PSD is observed within the beta frequency band, centered around \qty{21.6}{\Hz}, with no relevant infra-slow activity.
        \textbf{D-E)} A brief stimulus (\qty{2}{\ms}) was applied to the same nodes of the TUN and DEF configurations, to assess the spatio-temporal propagation of evoked activity. The post-stimulus power (see Sec.~\ref{subsubsec:mat_meth:metrics:time_freq_analysis}) of three subsets of nodes (specifically, the subset of stimulated nodes, a subset of far nodes connected to the stimulated nodes but in the same hemisphere where stimulation occurred, and the farthest set of nodes -- in terms of connectome tracts -- connected to the stimulated nodes), averaged across the same subset, displayed a consistent increase in all subsets in the TUN configuration with respect to DEF, together with a higher heterogeneity in the temporal evolution (compare D$_{\mathrm{I}}$ with E$_{\mathrm{I}}$). The extent of significant post-stimulus activity (see Sec.~\ref{subsubsec:mat_meth:metrics:complexity_metrics}), computed by summing the significant activations of each node between \num{10} and \qty{500}{\ms}, was mapped onto the brain map for TUN (D$_{\mathrm{II}}$) and DEF (E$_{\mathrm{II}}$). The panels indicate a more widespread propagation in TUN compared to DEF.
    }
\end{figure}

In this study, we take the first steps toward the calibration of a whole-brain model able to simultaneously reproduce key features of both spontaneous and evoked brain dynamics. To this end, we employ a recently developed whole-brain model implemented in TVB \citep{gaglioti2024investigating} and identify the most relevant parameters to be tuned.
Because the visualization and analysis tools included with TVB neither support systematic evaluation of metrics in response to changes in model parameters, nor allow quantitative comparison with experimental data, we set up an iterative approach (Fig.~\ref{fig:fig_1}A) that leverages analysis methods already implemented in Cobrawap (adding new ones suitably developed, when necessary).
Specifically, we identify the features that are relevant for describing both spontaneous and evoked activity of the whole-brain model: alpha-band oscillations, infra-slow fluctuations, and the spatio-temporal complexity of evoked responses. By recognizing and varying the related parameters, we can drive the model simulations into activity regimes that show more similarities to the heterogeneous, multiscale features of large-scale brain activity. The approach we propose aims at offering a first tackle to the above depicted process of tuning/calibration/validation of TVB-based brain models, a process that is not trivial, as discussed in~\citet{schirner2022brain, triebkorn2024fiftyshades}, especially when aiming at personalized models.
Our results show that an accurately tuned model can account for both the variability of resting-state activity and the complexity of stimulus-evoked responses, thus bridging a gap between traditionally separate modeling efforts for the two conditions. Moreover, we lay the groundwork for generalizable calibration/validation metrics that provide a robust foundation for future empirical testing and integration into automated workflows.

\section{Materials and Methods}
\label{sec:mat_meth}

In the following, we first detail the TVB implementation of the whole-brain model to be later tuned, followed by a description of the analysis steps used for measuring model features and performances.

\subsection{Model equations}
\label{subsec:mat_meth:model_eqs}

We use the Larter-Breakspear (LB) model \citep{larter1999coupled, breakspear2002nonlinear, breakspear2003modulation, breakspear2005dynamics}, a voltage-based phenomenological neural mass model commonly used to simulate whole-brain activity \citep{honey2007network, honey2008dynamical, alstott2009modeling, honey2009predicting, gollo2014frustrated, gollo2015dwelling, roberts2019metastable, endo2020evaluation, gaglioti2024investigating}, implemented within the TVB simulator.

Connectivity is implemented by means of a \num{998}-node bi-hemispherical connectome \citep{hagmann2008mapping}, where each node represents a source of signal reproducing the spontaneous dynamics of a neural mass model in the resting state.
The connectome specifies the strength of connections between nodes and the corresponding time delays, globally modulated by the coupling parameter $G$. Time delays between nodes can be globally rescaled by acting on the global conduction velocity parameter $c_v$.

The dynamics of a given node in the brain network is described by the following set of ordinary differential equations:

\begin{subequations}
    \label{eq:node_dynamics}
    \begin{align}
        \begin{split}
            \frac{1}{t_{\mathrm{scale}}} \, \frac{dV}{dt} &= -\left[g_{\mathrm{Ca}} + (1 - C) r_{\mathrm{NMDA}} \, a_{e \to e} \, Q_{\mathrm{V}} + C \, r_{\mathrm{NMDA}} \, a_{e \to e} \, Q_{\mathrm{V},\mathrm{network}}\right] \, m_{\mathrm{Ca}}(V - V_{\mathrm{Ca}}) \\
            & \qquad - \left[g_{\mathrm{Na}} \, m_{\mathrm{Na}} + (1 - C) \, a_{e \to e} \, Q_{\mathrm{V}} + C \, a_{e \to e} \, Q_{\mathrm{V},\mathrm{network}}\right] (V - V_{\mathrm{Na}}) \\
            & \qquad - g_{\mathrm{K}} \, W(V - V_{\mathrm{K}}) - g_{\mathrm{L}} (V - V_{\mathrm{L}}) - a_{i \to e} \, Z \, Q_{\mathrm{Z}} + a_{n \to e} \, I + \xi_{\mathrm{V}} \, ,
        \end{split}
        \label{eq:node_dynamics:V}
        \\
        \quad \nonumber
        \\
        \begin{split}
            \frac{1}{t_{\mathrm{scale}}} \, \frac{dZ}{dt} &= b \, \left(a_{n \to i} \, I + a_{i \to e} \, Q_{\mathrm{V}} \, V\right) + \xi_{\mathrm{Z}} \, ,
        \end{split}
        \label{eq:node_dynamics:Z}
        \\
        \quad \nonumber
        \\
        \begin{split}
            \frac{1}{t_{\mathrm{scale}}} \, \frac{dW}{dt} &= \frac{\phi \, (m_{\mathrm{K}} - W)}{\tau_{\mathrm{K}}} + \xi_{\mathrm{W}} \, ,
        \end{split}
        \label{eq:node_dynamics:W}
        \\
        \quad \nonumber
    \end{align}
\end{subequations}

\noindent
where $V$ and $Z$ are the mean membrane potential of the excitatory pyramidal neurons and inhibitory interneurons for that node, respectively, while $W$ corresponds to the average number of open potassium (K) channels. The term $t_{\mathrm{scale}}$ is an adimensional time-scaling factor that rescales the temporal dynamics of all state variables. The ratio of NMDA receptors to AMPA receptors is captured by $r_{\mathrm{NMDA}}$, and $a_{x \to y}$ defines the synaptic strength between source population $x$ (where $x$ is one of $e$: excitatory, $i$: inhibitory, or $n$: nonspecific input) and target population $y$ (where $y$ is $e$ or $i$). $I$ represents an external input current, modeling nonspecific excitatory drive, such as subcortical or thalamic inputs, that influences the activity of excitatory ($V$) and inhibitory ($Z$) populations. The dynamics of $Z$ and $W$ are governed by the constant rate terms~$b$, $\tau_\mathrm{K}$ and~$\phi$, which respectively represent a time constant for inhibitory activity ($Z$), a time constant for $W$ relaxation, and a temperature scaling factor for the evolution of $W$.

For each ion species, $g_{\mathrm{ion}}$ (where ion is one of Ca: calcium, Na: sodium, or K: potassium) represents the maximum ion conductance and $V_{\mathrm{ion}}$ represents its reversal potential. Similarly, $g_{\mathrm{L}}$ and $V_{\mathrm{L}}$ represent the maximum leakage conductance and reversal potential, respectively. The voltage-dependent fraction of open channels is governed by $m_{\mathrm{ion}}$, following a sigmoidal shape function for each node:

\begin{equation}
    m_{\mathrm{ion}}(t) = \frac{1}{2} \left[1 + \tanh\left(\frac{V(t) - T_{\mathrm{ion}}}{\delta_{\mathrm{ion}}}\right)\right] \, ,
\end{equation}

\noindent
where $T_{\mathrm{ion}}$ term represents the mean threshold membrane potential for a given ion channel population, while $\delta_{\mathrm{ion}}$ denotes its standard deviation. The mean firing rates of the excitatory and inhibitory node populations are governed by the voltage-dependent activation functions $Q_{\mathrm{V}}$ and $Q_{\mathrm{Z}}$, respectively, which are modeled as sigmoidal as well:

\begin{subequations}
    \label{eq:activation_functions}
    \begin{align}
        \begin{split}
            Q_{\mathrm{V}}(t) &= \frac{1}{2}\, Q_{\mathrm{V}_{\mathrm{max}}} \left[1 + \tanh\left(\frac{V(t) - V_{\mathrm{T}}}{\delta_{\mathrm{V}}}\right)\right] \, ,
        \end{split}
        \label{eq:activation_functions:QV}
        \\
        \quad \nonumber
        \\
        \begin{split}
            Q_{\mathrm{Z}}(t) &= \frac{1}{2}\, Q_{\mathrm{Z}_{\mathrm{max}}} \left[1 + \tanh\left(\frac{Z(t) - Z_{\mathrm{T}}}{\delta_{\mathrm{Z}}}\right)\right] \, ,
        \end{split}
        \label{eq:activation_functions:QZ}
        \\
        \quad \nonumber
    \end{align}
\end{subequations}

\noindent
where the terms $Q_{\mathrm{V}_{\mathrm{max}}}$ and $Q_{\mathrm{Z}_{\mathrm{max}}}$ are the maximum firing rates of the excitatory and inhibitory population, respectively. The thresholds for action potential generation for these populations are given by $V_{\mathrm{T}}$ and $Z_{\mathrm{T}}$, with corresponding standard deviations $\delta_{\mathrm{V}}$ and $\delta_{\mathrm{Z}}$, respectively. Finally, the input to a node $k$ coming from the rest of the network, $Q_{\mathrm{V},\mathrm{network}}^{(k)}$, is defined as:
\begin{equation}
    Q_{\mathrm{V},\mathrm{network}}^{(k)}(t) = G\sum_{\{j\}} u_{j \to k} \, Q^{(j)}_{\mathrm{V}}(t-\tau_{j \to k}) \, ,
\end{equation}
with the sum running over all the nodes $\{j\}$ connected to $k$ with connection weight $u_{j \to k}$, and $\tau_{j \to k}$ denoting the related input delay time. $G$ is the aforementioned global coupling that scales all the connection weights. The parameter $C$ in Eq.~(\ref{eq:node_dynamics}) varies between $0$ and $1$ and controls the balance between the strength of self-connections and the connections with the rest of the network. Finally, an additive Gaussian noise $\xi$ enters in Eq.~(\ref{eq:node_dynamics}) through the stochastic Heun integration method of TVB.

\subsection{Model configurations and parameters}
\label{subsec:mat_meth:model_configs_params}

We implement two parameter configurations for the LB model, with their values fully listed in Suppl. Tab.~\ref{tab:params}. The first configuration, referred to as the tuned configuration (TUN) from here on, is based on the work by~\citet{gaglioti2024investigating}, with some further refinements here. The second configuration, derived from the default parameters in TVB, follows the work of~\citet{alstott2009modeling} and will be referred to as the default configuration (DEF). The TUN configuration is capable of generating complex evoked patterns following external stimulation \citep{gaglioti2024investigating}, resembling empirical results from the healthy awake brain. In this study, we focus on resting-state activity and, to better align with empirical data, we adjust the timescale $t_{\mathrm{scale}}$ of the TUN configuration from \num{1.0} to \num{0.6} to reproduce the alpha-band location of the experimentally observed peak in the power spectrum of the global resting state activity \citep{berger1929uber, klimesch1999eeg, jensen2010shaping}. Notice that a time rescaling in Eq.~(\ref{eq:node_dynamics}) implies a coherent rescaling of the connectome conduction velocity $c_v$ and of the additive Gaussian noise $\xi$ in the TVB integrator (see Suppl. Tab.~\ref{tab:params}). Consistently, the TUN configuration exhibits a peak at \qty{11.7}{\Hz} in the resting condition (Fig.~\ref{fig:fig_1}B) and displays complex dynamics following perturbation (Fig.~\ref{fig:fig_1}D; see the following section for further details on the simulation of resting-state and evoked activity). The corresponding measures for the DEF configuration are depicted in Figs.~\ref{fig:fig_1}C and~\ref{fig:fig_1}E, respectively.

\subsection{Model activity simulation in TVB}
\label{subsec:mat_meth:model_sim_TVB}

Each simulation is run in TVB (RRID: SCR\_002249) with an integration time step of $\delta_{t}=\qty{0.1}{\ms}$ using the stochastic Heun integration method. A Gaussian white noise is added to the state variables to emulate an additional background input to the node (standard deviation $\sigma_{\xi} = 10^{-7}$ for the DEF configuration, then rescaled by $t_{\mathrm{scale}}$ for TUN). The signal is then resampled to a temporal resolution of \qty{1}{\ms} (i.e., averaging over \qty{1}{\ms} time windows with the \textit{temporal average monitor} of TVB).

For resting state, five minutes of continuous brain activity are simulated. Initial conditions are randomly assigned, with the values of the state variables ($V$, $Z$, $W$) drawn from a uniform distribution between \num{-0.1} and \num{0.1}. A conservative choice is made based on the observation that the initial transient phenomenon is concluded after roughly \qty{5}{\s}, hence an initial period of \qty{10}{\s} is excluded.

For evoked activity, \num{200} trials are simulated. A stimulus of \num{1} arbitrary unit (a.u.) in amplitude and \qty{2}{\ms} in duration is applied to the \num{27} nodes belonging to the right superior parietal (rSP) cortex region. Each trial begins with random initial conditions. The onset timing of the stimulus is randomized within a window between \qty{10}{\s} and \qty{12}{\s}, so again allowing for a long enough thermalization period after the random initialization. Finally, we retain \qty{400}{\ms} of pre-stimulus activity (baseline) and \qty{700}{\ms} of post-stimulus activity.

\subsection{Tuning TVB model parameters with Cobrawap}
\label{subsec:mat_meth:cobrawap}


We use the Cobrawap framework (RRID: SCR\_022966; \citep{cobrawap_docs, cobrawap_zenodo}) to implement the tuning process of TVB-simulated models. Cobrawap is an open-source Python-based modular analysis workflow that generates standardized quantitative descriptions of brain wave phenomena, so far used for heterogeneous murine datasets of both experimental~\citep{gutzen2024modular} and simulated~\citep{capone2023simulations} origin.
Cobrawap extensibility to a wider variety of use-cases -- among which, notably, the analysis of human data -- stems from its modular design: it consists of a series of workflow components implemented as scripts, each representing an elementary analysis or visualization operation; these \textit{blocks} are then organized into a series of sequential \textit{stages},
the selection and execution order of blocks in each stage being suitably configurable to fit the requirements of the input data and the analysis goal.

In this work we lay the foundations for the interoperability of TVB and Cobrawap -- acting on single blocks and on the Cobrawap structure itself -- as an integrated use-case for the complete calibration and the validation of TVB-simulated models.
Intrinsically, Cobrawap operates on matrices, reflecting the regular organization of sensors: each matrix element represents a \textit{channel} associated to the spatial position corresponding to an elementary source of experimental or simulated activity -- e.g., a camera pixel or an electrode -- sampled at discrete points in time.
To map TVB output to the Cobrawap framework, the individual neural mass of a particular node in the model is hence associated to a channel. In more detail, the 3D spatial arrangement of the N connectome nodes is reorganized as a 1D vector containing the same number of channels, arranged following the connectome ordering \citep{hagmann2008mapping}.
This simplified computational solution is sufficient for our analysis, since the focus is on node cross-correlations, for which only the distance among nodes in the connectome is relevant, together with TVB-encoded temporal delays.

New processing and visualization blocks are designed for this specific TVB-tuning use-case (see following section), improving and expanding what already available within the latest Cobrawap release \citep{cobrawap_zenodo}; these new features are currently implemented in a dedicated development software branch, and will be fully integrated in the official Cobrawap software release 0.3.0.


\subsection{Metrics employed for data analysis}
\label{subsec:mat_meth:metrics}

The following metrics are employed to comprehensively analyze the dynamics of simulated brain activity across both spontaneous and evoked conditions in the TUN and DEF configurations, unless otherwise specified.

\subsubsection{Events}
\label{subsubsec:mat_meth:metrics:events}

To characterize the node dynamics at the channel level, we identify the timestamps of transitions from low to high levels of activity in the signal evolution. Hereafter, we refer to such transitions as \textit{events}. Events are detected for each node by applying a threshold to the phase signal derived as the angle of the complex-valued analytic signal of the node (obtained via a Hilbert transform operation). In this study, a threshold of $-\pi/2$ is chosen, corresponding to the start of the upstroke. To ensure robustness, the algorithm selects only the time points where the threshold is crossed in an upward direction (from smaller to larger values), reaching a clearly identifiable peak ($\text{phase} = 0$) before the next threshold downward crossing. We obtain the \textit{average event rate} by counting the total number of events during the simulation divided by the time length of the simulation itself. Additionally, we define the time interval between events as \textit{inter-event-time} (IET), and calculate the related coefficient of variation $\mathrm{CV}_{\mathrm{IET}}$ (see details in the following).

\subsubsection{Power spectral density analysis}
\label{subsubsec:mat_meth:metrics:PSD}

According to the cortical organization principles summarized in \citet{larkum2013cellular}, superficial layers of the cortex are reached mainly by non-specific thalamic nuclei projections and inter-areal cortical association fibers, and the apical tuft of pyramidal neurons is functionally enriched for NMDA-dependent regenerative integration of these long-range paths. However, $Q_{\mathrm{V},\mathrm{network}}$ and $Q_{\mathrm{V}}$ excitatory contributions in Eq.~(\ref{eq:node_dynamics:V}) are not endowed with the slow injection times that are associated to NMDA receptors.
To create a better proxy for the EEG signal and investigate it in the frequency domain, we convolve $V$ with a bi-exponential kernel aimed to mimic synaptic currents. Also, this approach enables direct compatibility with the output of TVB simulations representing the average membrane potential of pyramidal population and not the impinging currents on that population. We select NMDA-like parameters to emulate the synaptic activity profile in the high-activation regime \citep[rise time constant $\tau_{\mathrm{rise}} = \qty{5}{\ms}$ and decay time constant $\tau_{\mathrm{decay}} = \qty{20}{\ms}$, following][]{gerstner2014neuronal}. Then, to obtain the synaptic activity profile of each cortical region, we average the convolved signal across all nodes corresponding to a given region (see Suppl. Tab.~\ref{tab:regions}), thus producing \num{66} regional synaptic activity profiles.
The power spectral density (PSD) of this synaptic activity proxy is then computed using Welch method with time segments of~\qty{70}{\s} and a \qty{50}{\percent} overlap.


As an additional measure, the PSD is computed on the hemispheric event data, obtained cumulating the number of events that occur at any node of each hemisphere using temporal bins of \qty{5}{\ms}; this results in two aggregated signals (one per hemisphere) with a sampling resolution of \qty{200}{\Hz}. To detect infra-slow frequencies, Welch method is applied with time segments of \qty{140}{\s} and a \qty{50}{\percent} overlap. On the resulting PSDs, the $1/f$ background scaling is estimated by fitting a linear function in log-log space to obtain the PSD slope, following~\citet{colombo2019spectral}.

\subsubsection{Time-frequency analysis}
\label{subsubsec:mat_meth:metrics:time_freq_analysis}

Time-frequency decomposition is performed using a continuous wavelet transform with complex Morlet wavelets, implemented via the \texttt{PyWavelets} package \citep{lee2019pywavelets}. In the resting state, scales (i.e., the temporal span of wavelets) are computed for frequencies around the main peak in the frequency spectrum (\qtyrange[range-phrase=--]{8}{12}{\Hz} for TUN and \qtyrange[range-phrase=--]{18}{22}{\Hz} for DEF configurations) in \qty{1}{\Hz} steps; for the evoked activity, scales range instead within \qtyrange[range-phrase=--]{8}{100}{\Hz}, still in \qty{1}{\Hz} steps, to account for a potentially wider evoked spectrum. Each wavelet is parameterized with a normalized bandwidth of \num{1} and a normalized center frequency of \num{1}; normalization depends on the sampling frequency (\qty{1}{\kHz}). The transformation is applied using a Fast Fourier Transform (FFT)-based method, yielding complex wavelet coefficients.

The time-frequency power representation is then obtained as the squared magnitude of the complex coefficients for each frequency, then averaged over the frequency range. Only for the analysis of the evoked activity, power is also averaged across all trials, to obtain the temporal evolution of broadband post-stimulus power. To account for baseline variability, trial-averaged power time-course $\mathcal{P}$ is normalized using a decibel (dB) transformation with respect to the pre-stimulus mean:
\begin{equation}
    \widetilde{\mathcal{P}} \equiv 10 \log_{10}{\Bigl(\mathcal{P}_{\mathrm{post}}/\langle \mathcal{P}_{\mathrm{pre}}\rangle\Bigr)} \, ,
\end{equation}
where $\mathcal{P}_{\mathrm{post}}$ represents the post-stimulus power at each time point, while $\langle\mathcal{P}_{\mathrm{pre}}\rangle$ is the time average from \qty{-350}{\ms} to \qty{-50}{\ms} with respect to the stimulus onset.

To assess long-range temporal correlations in resting-state neural dynamics, we apply the Detrended Fluctuation Analysis (DFA) to the amplitude envelope from wavelet transform (i.e., the frequency average of complex coefficients magnitude). DFA quantifies long-range temporal correlations -- a hallmark of critical dynamics \citep{linkenkaerhansen2001longrange} -- by measuring how fluctuations persist over increasing time window sizes. We use the implementation provided in the \texttt{neurodsp} package \citep{cole2019neurodsp}, which first detrends the signal by removing its mean and computes the cumulative sum. The resulting signal is then divided into equal-sized windows, and within each, a linear trend is fitted. The fluctuation is computed as the mean squared deviation from this trend, and the procedure is repeated over multiple window sizes to estimate the scaling exponent. In our analysis, we use \num{50} window sizes, logarithmically spaced between \num{0.5} and \num{50} seconds.

\subsubsection{Correlation functions}
\label{subsubsec:mat_meth:metrics:corr_funcs}

The auto-correlation function quantifies the temporal correlation of a signal with itself across varying time lags. In this study, we apply it to channel signals from both TUN and DEF configurations, considering lags up to \qty{\pm 120}{\ms}, in steps of \qty{1}{\ms}. The analysis focuses on identifying the largest auto-correlation peak at $\text{lags} \neq 0$, and the corresponding \textit{optimal} value of the lag is recorded for further analysis.
Analogously, the cross-correlation function is employed to quantify the similarity between signals originating from different brain nodes across varying time lags, also in this case up to \qty{\pm 120}{\ms} in steps of \qty{1}{\ms}.
This analysis allows us to extract two key measures for each ordered pair of nodes: functional connectivity $\mathbb{FC}$, defined as the peak value of the cross-correlation function, and functional lag $\mathbb{FL}$, corresponding to the lag (in \unit{\ms}) at which the peak occurred, where the sign of the lag reflects the direction of the interaction (e.g., a negative lag between nodes $i$ and $j$ would indicate a directional relation from $i$ to $j$).
In general, this approach results in non-symmetric $\mathbb{FC}$ and $\mathbb{FL}$ matrices, thereby capturing the directionality of interactions between brain nodes. Hence, we quantify such asymmetry as $\mathbb{FC}_{\mathrm{asym}} \equiv \mathbb{FC} - \mathbb{FC}^{\intercal}$ and $\mathbb{FL}_{\mathrm{asym}} \equiv \mathbb{FL} - \mathbb{FL}^{\intercal}$, where ${(\cdot)}^{\intercal}$ denotes the matrix transpose. In particular, the matrix $\mathbb{FC}_{\mathrm{asym}}$ captures directional differences in communication, with positive values $\mathbb{FC}_{\mathrm{asym},ij} > 0$ indicating that node $i$ exerts a stronger influence on node $j$ than vice versa, and negative values $\mathbb{FC}_{\mathrm{asym},ij} < 0$ indicating the opposite.
To derive a nodal asymmetry score, we finally compute the sum of asymmetry values across each row of $\mathbb{FC}_{\mathrm{asym}}$, thus projecting the 2D representation of the matrix into 1D vectors, and yielding a single asymmetry score per node:
\begin{equation}
    \mathbb{A}_{\mathrm{i}} = \sum_{j} \mathbb{FC}_{\mathrm{asym},ij} \, ,
\end{equation}
where $\mathbb{A}_{i}$ represents the net asymmetry of node $i$, with positive values indicating a global leading role (\textit{parent}) in the network, and negative values indicating nodes behaving as ``followers'' (\textit{children}).

\subsubsection{Complexity metrics: PCI and functional complexity}
\label{subsubsec:mat_meth:metrics:complexity_metrics}

The \textit{Perturbational Complexity Index} (PCI), originally introduced by~\citet{casali2013theoretically}, provides a quantitative measure of the inherent complexity of spatio-temporal EEG activity patterns evoked by Transcranial Magnetic Stimulation (TMS). In this study, we adopt a similar framework, leveraging evoked whole-brain activity simulated using TVB as a proxy of the signal reconstructed at the ``sources''.
To identify the significant spatio-temporal patterns of stimulus-evoked responses, we apply the non-parametric bootstrap-based statistical analysis of~\citet{casali2013theoretically} on the voltage signal of each node. The procedure, also known as \textit{maximal statistic} \citep{nichols2001nonparametric}, is organized as follows. First, the evoked activity is rescaled with respect to the pre-stimuls baseline by subtracting the mean and dividing by the standard deviation of the pre-stimulus period. Second, for each node, pre-stimulus activities are randomly sampled with replacement (\textit{bootstrapped}) from the original distribution across trials and pre-stimulus time samples. Third, for each node and time point, the mean across trials of the bootstrapped values are computed (\textit{Jboot}). Fourth, for each pre-stimulus time point, the maximum absolute value of Jboot across nodes is obtained. This process (steps 2, 3 and~4) is repeated \num{500} times to build a null distribution from which the ($1-\alpha$) percentile is used to define a significance threshold (we use $\alpha=0.05$). Finally, a binary spatio-temporal mask is obtained by marking samples exceeding this threshold as significant, as in~\citet{casali2013theoretically}. Eventually, PCI is computed by quantifying the Lempel-Ziv complexity \citep{lempel1976complexity} of the binarized matrix representing significant activations, then normalized by the source entropy.

In addition, in order to assess the richness of interactions within functional connectivity matrix~$\mathbb{FC}$, \textit{functional complexity} can be computed, which quantifies the balance between functional integration and functional segregation \citep{zamoralopez2016functional}. A low functional complexity is characterized by narrow distributions of values in the matrix, indicating either near-total statistical independence ($\langle \mathbb{FC} \rangle \approx 0$, i.e. total functional segregation) or global synchrony ($\langle \mathbb{FC} \rangle \approx 1$, i.e. total functional integration). In contrast, complex interactions arise when the collective activity is characterized by intermediate states yielding to a broad distribution of the $\mathbb{FC}$ values. Following this methodology, functional complexity is computed as the sum of the absolute differences over the $m$ bins between the distribution of~$\mathbb{FC}$ values and the uniform distribution over the same range, as described in~\citet{zamoralopez2016functional}.

\subsubsection{Statistical metrics}
\label{subsubsec:mat_meth:metrics:stats_metrics}

Standard statistical metrics are employed in this study, including the standard deviation (SD), representing the dispersion around the mean, and the inter-quartile range (IQR), describing the spread of the middle \qty{50}{\percent} of the data. The coefficient of variation ($\mathrm{CV}$) is used to assess relative variability (defined as the ratio of the standard deviation to the mean), while the median is used as a robust measure of central tendency.

\section{Results}
\label{sec:results}

We simulate a whole-brain model via TVB, relying on a Larter-Breakspear neural mass model with a \num{998}-node human connectome (see Secs.~\ref{subsec:mat_meth:model_eqs}-\ref{subsec:mat_meth:model_sim_TVB}), considering both the spontaneous activity in the resting state and the \textit{evoked} one (i.e., in response to an external perturbation). To assess the robustness of the approach, we also run a subset of the analyses on a smaller scale \num{74}-node human connectome, as detailed in Supplementary Methods. We leverage Cobrawap (see Sec.~\ref{subsec:mat_meth:cobrawap}) to analyze the simulation output and tune the model parameters so to match a set of quantitative metrics based on relevant biological features (see Sec.~\ref{subsec:mat_meth:metrics}) not expressed by the initial default (DEF) model configuration. Specifically: in response to external perturbations, the tuned (TUN) model generates non-stereotyped, complex spatio-temporal activity, as quantified by the perturbational complexity index; in spontaneous activity, it displays robust alpha-band oscillations, infra-slow rhythms, scale-free characteristics, greater spatio-temporal heterogeneity, and asymmetric functional connectivity.

We focus the tuning procedure essentially on the connectome global coupling $G$, the connectome conduction velocity $c_v$, the overall timescale factor $t_{\mathrm{scale}}$, and a subset of parameters in the equations ruling the single-node dynamics. Refer to Sec.~\ref{sec:mat_meth} for details about their definition, and to Suppl. Tab.~\ref{tab:params} for their values in both DEF and TUN configurations.

The relevance of properly tuning the global coupling $G$ (here increased from \num{1} to \num{3}) to induce complex, non-stereotyped dynamics in this model -- resembling the experimentally observed one -- has been preliminarily demonstrated in~\citet{gaglioti2024investigating}, where the target is to match functional and structural connectivity, along the lines proposed by~\citet{deco2014identification}. In the same study~\citep{gaglioti2024investigating}, the tuning of node parameters is instrumental in obtaining evoked responses comparable to TMS-EEG experiments, as also discussed hereafter.
In addition, here we focus on reproducing the previously listed biological features relevant in the spontaneous resting-state activity, again acting on single-node equations parameters. Among them, concerning the expression of proper rhythms, we identify as essential parameters to be tuned the overall timescale factor $t_{\mathrm{scale}}$ (moved from \num{1.0} to \num{0.6}) and the conduction velocity $c_v$ (rescaled for coherence by the same amount of $t_{\mathrm{scale}}$; see Suppl. Tab.~\ref{tab:params}).

In what follows, we quantitatively detail the different behavior of the two model configurations, highlighting the importance of accurately tuning the model parameters for moving from a stereotyped simulation toward a more biologically plausible one.

\subsection{Characterizing spontaneous dynamics: divergent temporal and spectral patterns in tuned vs. default configurations}
\label{subsec:results:spont_dynamics_cobrawap}

We start by examining the membrane potential traces recorded in the resting-state condition, shown for three representative nodes (picked from FUS, PREC, and RMF regions of the right hemisphere, respectively; see Suppl. Tab.~\ref{tab:regions}) in TUN (Fig.~\ref{fig:fig_2}A) and DEF (Fig.~\ref{fig:fig_2}C) configurations; the corresponding amplitude distributions are shown in Figs.~\ref{fig:fig_2}B and~\ref{fig:fig_2}D, respectively. Qualitatively, the TUN configuration exhibits more diverse activity patterns, with some nodes also displaying bursting dynamics (e.g., middle trace in Fig.~\ref{fig:fig_2}A).

\begin{figure}[!b]
    \centering
    \includegraphics[width=\linewidth]{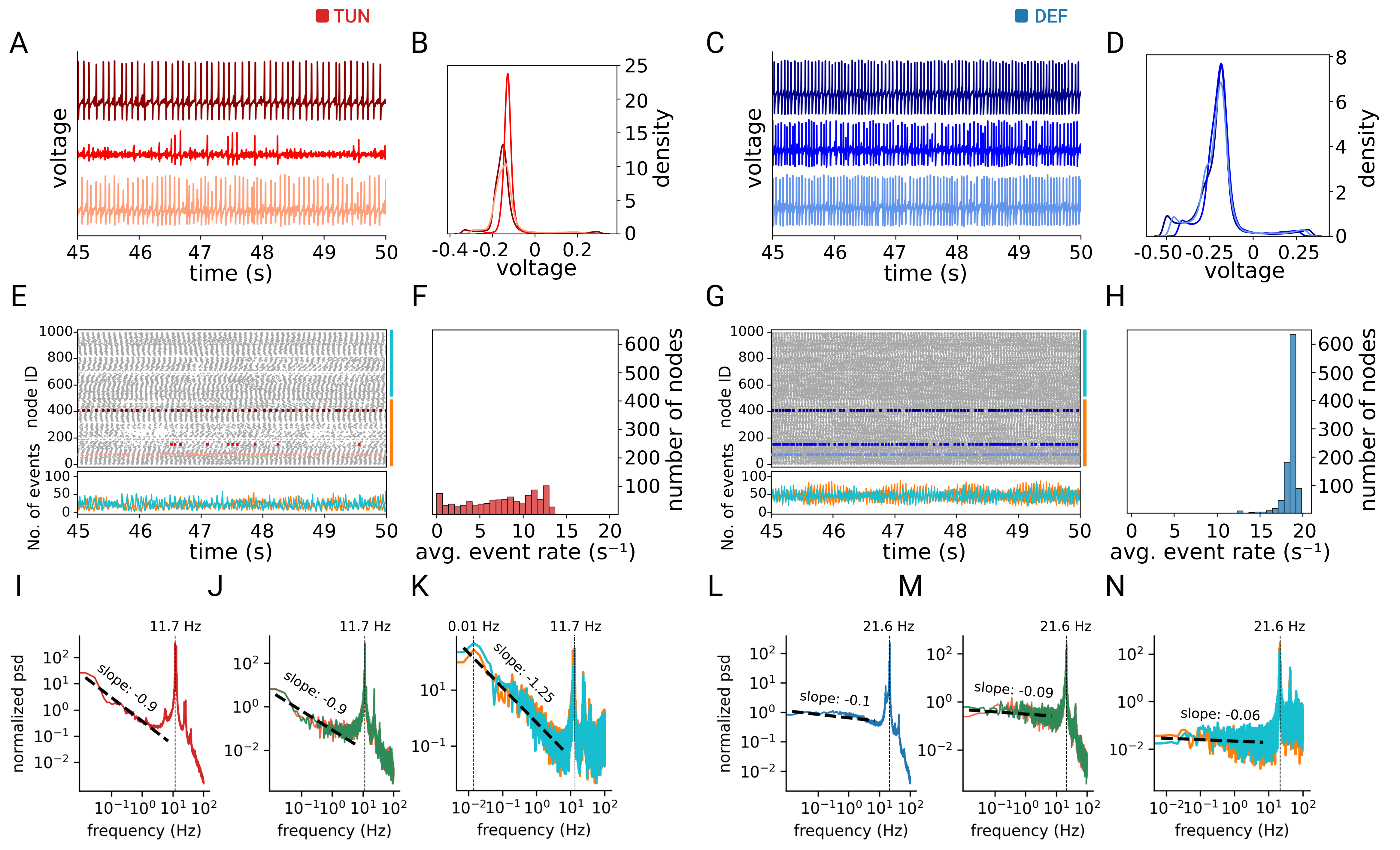}
    \caption{
        \textbf{Comparison of spontaneous dynamics in TUN and DEF configurations: time-domain and frequency-domain analyses.}\\
        \textbf{A)} Voltage traces from three representative nodes from right-hemisphere FUS, PREC, and RMF regions (brown, red, and salmon lines, respectively) in the TUN configuration.
        \textbf{B)} The probability density functions of voltage distributions from nodes in A). 
        \textbf{C-D)} Same as A) and B), respectively, for the DEF configuration.
        \textbf{E)} Rasterplots of upgoing transition events for each node (gray, top panel), alongside the cumulated number of detected events on the right and left hemisphere (orange and cyan, respectively, bottom panel), in the TUN configuration. Colored events in the upper panel are relative to the same sample traces shown in A).
        \textbf{F)} Histograms of the average event rate (red filled bars) across all nodes in TUN.
        \textbf{G-H)} Same as E) and F), respectively, for the DEF configuration.
        \textbf{I)} Average PSD of synaptic activity proxy across all regions
    }
    \label{fig:fig_2}
\end{figure}

\begin{figure}[!t]
    \centering
    \ContinuedFloat
    \caption*{
        in TUN, highlighting a peak frequency at \qty{11.7}{\Hz} (black vertical dashed line).
        \textbf{J)} PSD of synaptic activity proxy of left LOCC (green line) and right LOCC (red line) in TUN, with a peak frequency at \qty{11.7}{\Hz}. The synaptic activity was averaged over the nodes of right and left LOCC before computing the PSD.
        \textbf{K)} PSD of the cumulated event signals from both hemispheres (right hemisphere: orange line; left hemisphere: cyan line), showing a low-frequency content ($\sim$ \qty{0.01}{\Hz}) in TUN. The black dashed oblique line represents the spectral slope, obtained by fitting a line to the average PSD in log-log space, in the frequency range \qtyrange[range-phrase=--]{0.01}{7}{\Hz}.
        \textbf{L-N)} Same as I), J) and K), respectively, for the DEF configuration; notice the peak frequency at \qty{21.6}{\Hz}, and the absence of both the low-frequency content and the $1/f$ trend in the PSD of the default configuration.
    }
\end{figure}

To identify the events of our interest, i.e. changes in the neural activity, we analyze the Hilbert phase of the membrane potential and apply a threshold of $-\pi/2$, following the methodology outlined by~\citet{gutzen2024modular} (see Sec.~\ref{subsubsec:mat_meth:metrics:events}). Figs.~\ref{fig:fig_2}E and~\ref{fig:fig_2}G show the rasterplots of upgoing events for the two configurations, whereas Figs.~\ref{fig:fig_2}F and~\ref{fig:fig_2}H illustrate the histograms of the average event rate across nodes (see Sec.~\ref{subsubsec:mat_meth:metrics:events}); cyan and orange bars on the left side of the two rastergrams identify nodes belonging to the two hemispheres (left and right, respectively). The same color coding is used in the bottom part of Figs.~\ref{fig:fig_2}E and~\ref{fig:fig_2}G, where cyan and orange traces represent the cumulated event signal of each hemisphere (see Sec.~\ref{subsubsec:mat_meth:metrics:PSD}). 
A clear distinction emerges in the distribution of the average event rate across nodes: TUN shows a more heterogeneous pattern, compared to DEF (mean and standard deviation across channels: \qty{9.60 \pm 7.41}{\Hz} for TUN, \qty{18.56 \pm 0.91}{\Hz} for DEF).

The synaptic activity proxy (see Sec.~\ref{subsubsec:mat_meth:metrics:PSD}) is derived from the voltage traces to approximate post-synaptic currents and obtain synaptic activity profiles for each cortical region (exemplary traces shown in Fig.~\ref{fig:fig_1}B-C). We compute both the average PSD across all cortical regions (Fig.~\ref{fig:fig_2}I for TUN, and Fig.~\ref{fig:fig_2}L for DEF) and the PSD for specific regions of interest (Fig.~\ref{fig:fig_2}J for TUN, and Fig.~\ref{fig:fig_2}M for DEF) -- here, the left and right lateral occipital cortex (LOCC; see Suppl. Tab.~\ref{tab:regions}). Additionally, to study the collective dynamics of the network, we compute the PSD from the cumulated event signal of the two hemispheres (Fig.~\ref{fig:fig_2}K for TUN, and Fig.~\ref{fig:fig_2}N for DEF; see Sec.~\ref{subsubsec:mat_meth:metrics:PSD}), derived from the rasterplots of events. By construction of the configurations, DEF displays a dominant frequency peak at \qty{21.6}{\Hz}, whereas TUN shows a peak in the alpha band at \qty{11.7}{\Hz}. This shift is primarily driven by the change in the timescale parameter $t_{\mathrm{scale}}$ with respect to default setting (from \num{1.0} to \num{0.6}; see Suppl. Tab.~\ref{tab:params} and Sec.~\ref{sec:mat_meth}). Furthermore, TUN reveals a distinct peak at \qty{0.01}{\Hz} in the PSD of the cumulated event signal -- absent in DEF -- highlighting infra-slow network fluctuations; in contrast to the dominant frequency, this effect cannot be readily attributed to the tuning of a single parameter. Notably, a $1/f$ trend is also observed in TUN, but is absent in DEF, suggesting scale-free dynamics in the former configuration only.

The spectral features of the TUN configuration, namely alpha-band power and $1/f$ scaling, also show a closer correspondence to empirical EEG recordings obtained during resting state with eyes closed (Suppl. Fig.~\ref{fig:fig_S1}), further supporting its biological plausibility.

Overall, these findings indicate that the TUN configuration supports a richer repertoire of dynamical activity across nodes, characterized by more heterogeneous average event rates, prominent infra-slow network fluctuations, and a scale-free $1/f$ spectral profile -- hallmarks of dynamics closer to a critical regime \citep{chialvo2010emergent, palva2018roles}.

\subsection{Signal periodicity analysis reveals spatially organized spectral features in the tuned configuration}
\label{subsec:results:signal_periodicity}

To better understand the apparent mismatch in TUN between the heterogeneous event rates across nodes (Fig.~\ref{fig:fig_2}F) and the spectral peak at \qty{\sim 12}{\Hz} that dominates at the global scale (Figs.~\ref{fig:fig_2}I-K), we turn to auto-correlation analysis of the spontaneous voltage activity to capture the degree of temporal regularity of oscillations at each node. We therefore compute the auto-correlation function for each node in both TUN and DEF configurations.

\begin{figure}[!t]
    \centering
    \includegraphics[width=\linewidth]{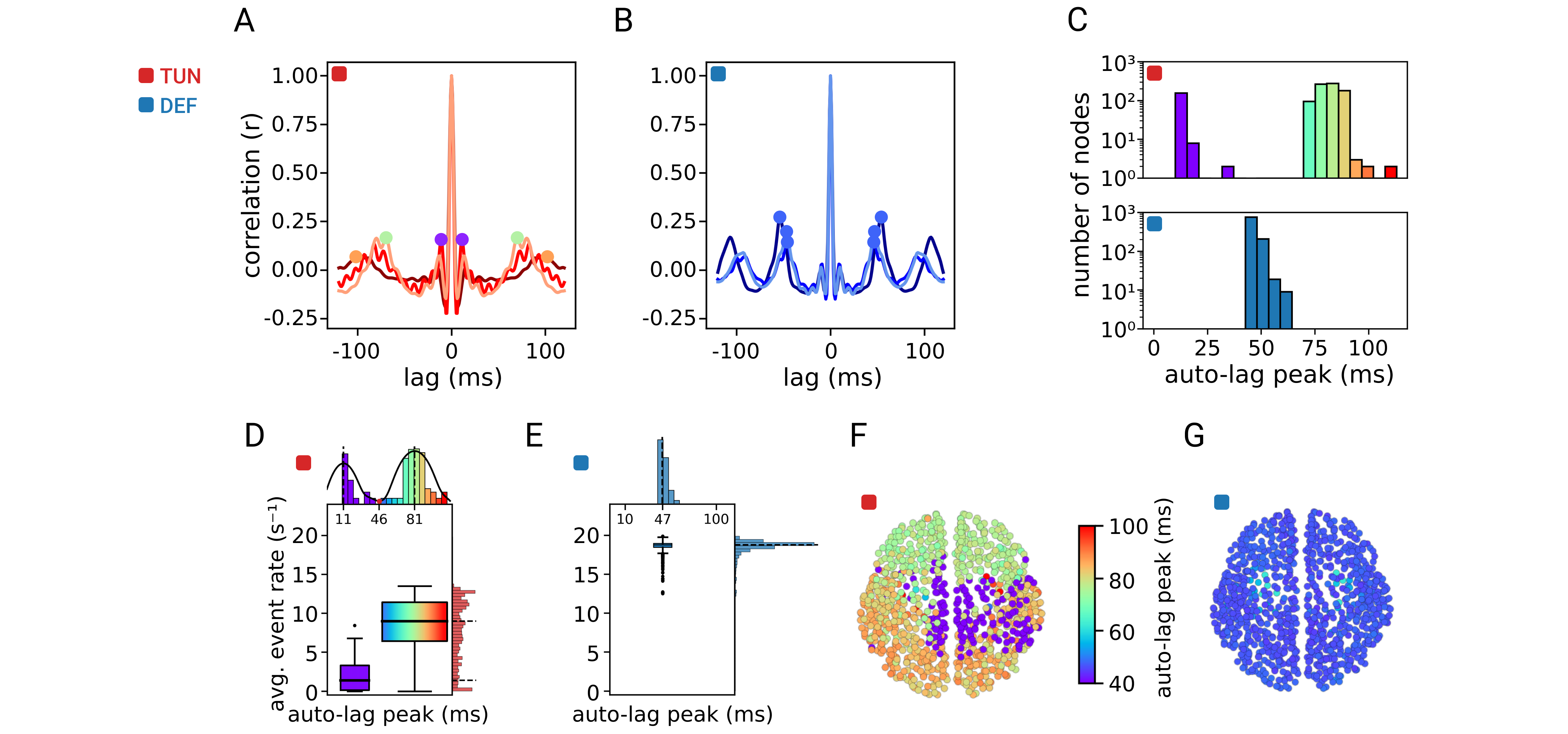}
    \caption{
        \textbf{Auto-correlation analysis and spatial distribution of periodicity in TUN and DEF.}\\
        \textbf{A-B)} Auto-correlation functions of the spontaneous voltage signal from three representative nodes (same as in Fig.~\ref{fig:fig_2}) for TUN (red palette) and DEF (blue palette). The largest peak at $\text{lag} \ne 0$ is marked by coloured dots.
        \textbf{C)} Distribution of the latencies of the dominant auto-correlation peaks ($\text{lag} > 0$) across brain nodes, revealing a bimodal distribution in TUN (top), while DEF (bottom) exhibits a narrow distribution centered around \qty{50}{\ms}.
        \textbf{D)} Relationship between signal periodicity and average event rate in TUN. Boxplots depict the average event rate for nodes grouped by short ($ < \qty{46}{\ms}$) and long ($> \qty{46}{\ms}$) auto-correlation peaks. The nodes exhibiting periodicity in the first mode (i.e. $< \qty{46}{\ms}$) are \qty{17}{\percent} of the total. The vertical and horizontal black dashed lines on the histogram mark the median of the distributions.
        \textbf{E)} Same as D), but for DEF. Here nodes are considered all together, not split into subgroups according to their auto-lag peaks.
        \textbf{F-G)} Spatial distribution of peak latencies mapped onto brain nodes for TUN and DEF configurations. TUN displays structured clustering, with faster periodicities in medial-central regions and a gradient between frontal and posterior regions. DEF exhibits a more homogeneous distribution of periodicities across the brain.
     }
    \label{fig:fig_3}
\end{figure}

These functions are illustrated in Figs.~\ref{fig:fig_3}A and~\ref{fig:fig_3}B for three different nodes (the same as in Fig.~\ref{fig:fig_2}, from right FUS, PREC, and RMF regions). By examining the largest auto-correlation peak at $\text{lag} \ne 0$, we observe a median periodicity of \qty{80}{\ms} (\qty{12.5}{\Hz}) in TUN and \qty{47}{\ms} (\qty{21.3}{\Hz}) in DEF. This periodicity aligns with the PSD results in Figs.~\ref{fig:fig_2}I and~\ref{fig:fig_2}L, where a peak around \qty{12}{\Hz} is observed for TUN, and around \qty{22}{\Hz} for DEF.

We therefore hypothesize that the median periodicity of \qty{80}{\ms} and \qty{47}{\ms} observed in TUN and DEF, respectively, is driven by the temporal arrangement of events in the two configurations. Supporting this hypothesis, the node-averaged PSD of the events (see Suppl. Fig.~\ref{fig:fig_S2}) reveals peaks at approximately \qty{12}{\Hz} in TUN and \qty{22}{\Hz} in DEF. We interpret this as a signature of the network periodicity (as also suggested by the PSD of the hemispheric signal in Figs.~\ref{fig:fig_2}K and~\ref{fig:fig_2}N), which we suggest is predictably influenced by global model parameters, such as conduction velocity $c_v$ and global coupling $G$, as detailed in Suppl. Fig.~\ref{fig:fig_S3}.

In addition to differences in the median peak, TUN exhibits also shorter peak latencies of approximately \qty{11}{\ms}, as shown for a node of the right PREC region in Fig.~\ref{fig:fig_3}A (red line and purple circles).

The distribution across nodes of the latencies of the dominant auto-correlogram peaks (Fig.~\ref{fig:fig_3}C, with $\text{lag} > 0$) highlights a clear difference between the TUN (Fig.~\ref{fig:fig_3}C, top panel) and the DEF (Fig.~\ref{fig:fig_3}C, bottom panel) configurations. In DEF the distribution is centered around \qty{50}{\ms} (median = \qty{47}{\ms}) with low variability (SD = \qty{2.15}{\ms}), while in TUN a bimodal distribution emerges, with a larger standard deviation (SD = \qty{26.48}{\ms}) driven by the presence of two distinct modes centered around \qty{11}{\ms} and \qty{81}{\ms}. The underlying cause for this bimodality is the relative difference in the strength of the \qty{\sim 80}{\ms} and \qty{\sim 11}{\ms} auto-correlogram peaks, varying node by node.

Since the auto-lag at \qty{\sim 80}{\ms} is mainly driven by the model events, we hypothesize that nodes expressing a higher strength of \qty{\sim 11}{\ms} peaks are those with a low average event rate (i.e., lower number of events), reflecting sub-threshold activity typical of nodes with irregular and bursty dynamics. An example is shown in the middle trace of Fig.~\ref{fig:fig_2}A, corresponding to the node in the right PREC region with a peak at \qty{11}{\ms} in Fig.~\ref{fig:fig_3}A.

To test this, the two modes in the TUN auto-lag peak distribution are split considering the minimum of its Kernel Density Estimate function (found at \qty{46}{\ms}); this allows to separate the nodes in two groups, depending on the dominant peak in the auto-correlogram. Relating the nodes of either group to their average event rate (Fig.~\ref{fig:fig_3}D), we observe that the average event rate in TUN is lower for nodes in the group with short lags (median=\qty{1.38}{\Hz}; mean=\qty{1.92}{\Hz}; SD=\qty{1.93}{\Hz}) than for nodes in the group with long lags (median=\qty{9.00}{\Hz}; mean=\qty{8.66}{\Hz}; SD=\qty{3.35}{\Hz}). We repeat the same analysis for DEF in Fig.~\ref{fig:fig_3}E: the unimodality of DEF in the distribution of the latency is reflected by an analogous unimodality in the distribution of the average event rate (median=\qty{18.78}{\Hz}; mean=\qty{18.56}{\Hz}; SD=\qty{0.91}{\Hz}). In summary, we interpret these findings such that the broad distribution of events and auto-lag peaks in the TUN configuration (vertical and horizontal histograms in Fig.~\ref{fig:fig_3}D, respectively) is due to a superposition of a regular alpha-band activity and a component at \qty{11}{\ms}, driven by the irregular, bursting activity of nodes exhibiting more sub-threshold dynamics (i.e., lower number of events). In comparison, the DEF configuration shows a stereotyped beta oscillation that dominates the lags (Fig.~\ref{fig:fig_3}E).

To visualize the spatial organization of these differences, we map the peak latencies onto a brain map (Figs.~\ref{fig:fig_3}F and~\ref{fig:fig_3}G), finding that TUN exhibits a more spatially organized clustering structure compared to DEF. Indeed, in TUN (Fig.~\ref{fig:fig_3}F) nodes with peak latencies around \qty{11}{\ms} are clustered within central-parietal areas. Additionally, considering the second mode after \qty{46}{\ms}, shorter peak latencies (corresponding to faster oscillations in the auto-correlation trace) are observed in frontal compared to posterior regions (Fig.~\ref{fig:fig_3}F), an emergent feature of the model configuration that trends toward experimental findings \citep{rosanova2009natural, frauscher2018atlas, capilla2022natural}. Conversely, DEF displays a homogeneous spatial distribution, corresponding to more stereotypical simulated dynamics (Fig.~\ref{fig:fig_3}G).

Together, these results demonstrate that the tuned configuration generates richer spatio-temporal dynamics characterized by \textit{i)} a non-trivial spatial organization of the rhythms (with both frontal-posterior latency gradients and distinct central-parietal clusters showing shorter latencies and reduced average event rates), and \textit{ii)} an enhanced overall heterogeneity -- all contrasting sharply with the homogeneous, stereotyped activity patterns in the default configuration of the model.

\subsection{Spatio-temporal irregularity and scale-free dynamics in the tuned configuration}
\label{subsec:results:spatiotemporal_irregularity}

\begin{figure}[!t]
    \centering
    \includegraphics[width=\linewidth]{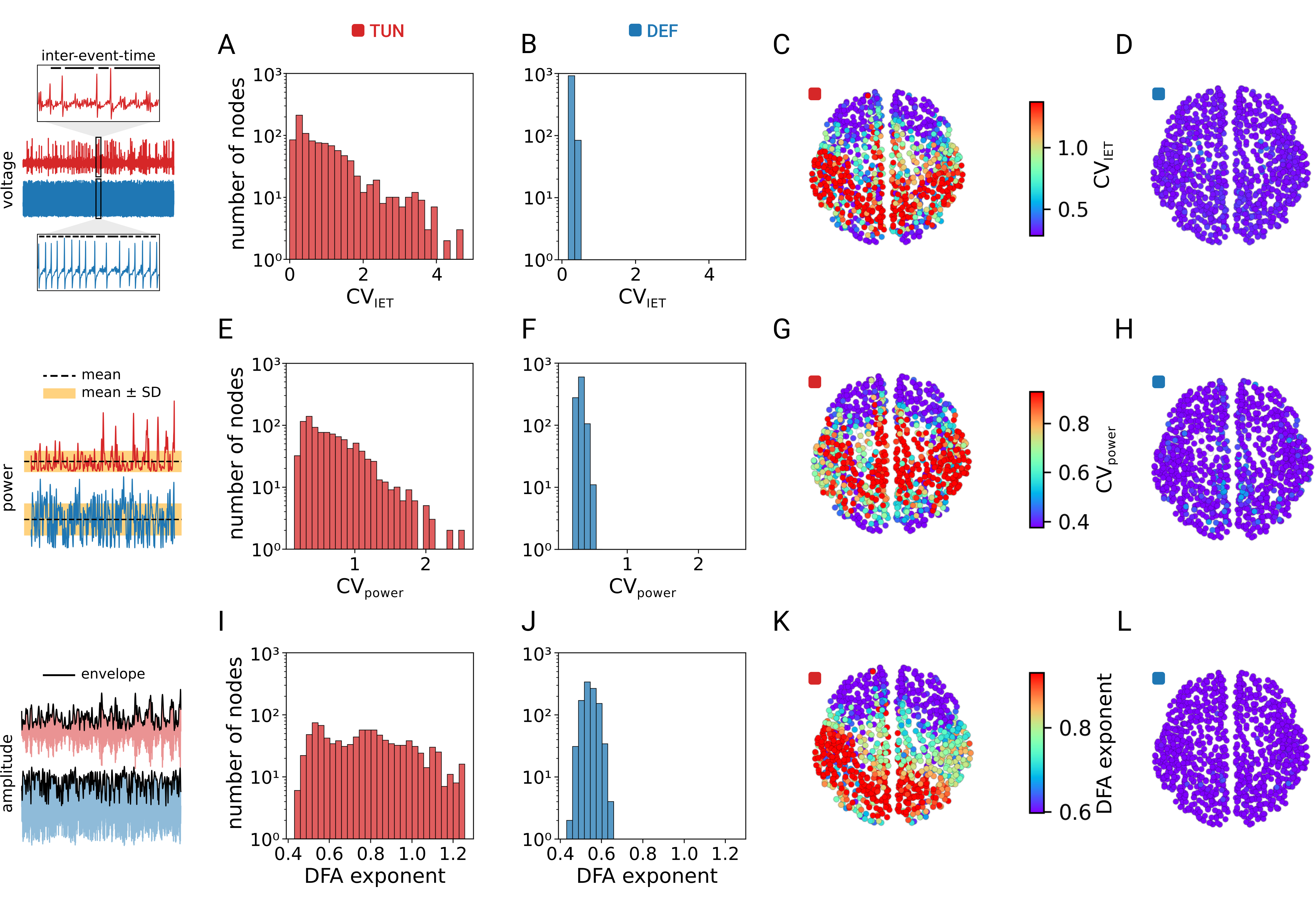}
    \caption{
        \textbf{Temporal and spatial variability of signal fluctuations in TUN and DEF.}\\
        Signal processing steps are illustrated in the leftmost panels of each row: Top: Example traces show voltage fluctuations and inter-event-time (IET) used to calculate the $\mathrm{CV}_{\mathrm{IET}}$. Middle: Example power time courses used to compute the $\mathrm{CV}_{\mathrm{power}}$, with shaded regions indicating mean $\pm$ standard deviation. Bottom: Example amplitude envelope traces used to estimate the DFA exponent across nodes.
        \textbf{A-B)} Histograms showing the coefficient of variation of inter-event-time ($\mathrm{CV}_{\mathrm{IET}}$) across nodes for TUN and DEF configurations.
        \textbf{C-D)} Spatial distribution of the $\mathrm{CV}_{\mathrm{IET}}$ mapped onto brain nodes for TUN and DEF configurations.
        \textbf{E-F)} Histograms showing the coefficient of variation of the power time-courses ($\mathrm{CV}_{\mathrm{power}}$) across nodes, computed via Morlet wavelet convolution for TUN and DEF configurations.
        Power was obtained by summing within the \qtyrange[range-phrase=--]{8}{12}{\Hz} range for TUN and the \qtyrange[range-phrase=--]{18}{22}{\Hz} range for DEF.
        \textbf{G-H)} Brain maps displaying the spatial distribution of $\mathrm{CV}_{\mathrm{power}}$ in TUN and DEF configurations.
        \textbf{I-J)} Histograms showing the DFA exponent computed for each node on the amplitude envelope in the \qtyrange[range-phrase=--]{8}{12}{\Hz} range for TUN and the \qtyrange[range-phrase=--]{18}{22}{\Hz} range for DEF.
        \textbf{K-L)} Brain maps displaying the spatial distribution of DFA exponents for TUN and DEF configurations.
     }
    \label{fig:fig_4}
\end{figure}

The results presented in Fig.~\ref{fig:fig_3} demonstrate that the TUN configuration exhibits spectral heterogeneity which is reflected in a distinctive spatial topography of auto-correlation peak latencies. To explore how this heterogeneity manifests in the temporal domain, we next compute the coefficient of variation of the inter-event-time ($\mathrm{CV}_{\mathrm{IET}}$; see Sec.~\ref{subsubsec:mat_meth:metrics:events}) across nodes. Figs.~\ref{fig:fig_4}A and~\ref{fig:fig_4}B show the distribution of $\mathrm{CV}_{\mathrm{IET}}$ for TUN and DEF, respectively, while Figs.~\ref{fig:fig_4}C and~\ref{fig:fig_4}D map these values onto the brain. Once again, we observe a more pronounced heterogeneity in TUN, which is also reflected in a non-trivial spatial organization, in contrast with DEF. Specifically, TUN exhibits more irregular (i.e., higher $\mathrm{CV}_{\mathrm{IET}}$ overall, particularly in central-parietal areas) and spatially heterogeneous events, while DEF shows more regular (i.e., lower $\mathrm{CV}_{\mathrm{IET}}$) and spatially homogeneous event patterns.

Building on this observation of irregularity and heterogeneity in event timing, a complementary and related aspect to consider is the temporal fluctuation of the neural signals themselves. To examine these dynamics more closely, we perform a time-frequency decomposition using Morlet wavelet convolution to extract the power time-course within the frequency band associated with events, specifically, the \qtyrange[range-phrase=--]{8}{12}{\Hz} range for TUN and the \qtyrange[range-phrase=--]{18}{22}{\Hz} range for DEF, corresponding to the respective dominant frequencies in their PSDs. The power time-courses (see Sec.~\ref{subsubsec:mat_meth:metrics:time_freq_analysis}) of a representative node in TUN and DEF are reported in Suppl. Fig.~\ref{fig:fig_S4}A. To quantify power fluctuations over time, we compute their coefficient of variation ($\mathrm{CV}_{\mathrm{power}}$) across nodes (Figs.~\ref{fig:fig_4}E and~\ref{fig:fig_4}F) and map it onto brain maps (Figs.~\ref{fig:fig_4}G and~\ref{fig:fig_4}H). Also in this case, power exhibits a high temporal and spatial regularity for DEF configuration, resulting in a more homogeneous brain map with low and quite concentrated $\mathrm{CV}_{\mathrm{power}}$ values. Conversely, we observe a heterogeneous distribution for TUN, with central areas displaying high $\mathrm{CV}_{\mathrm{power}}$ values, indicative of pronounced power fluctuations. Interestingly, the average PSD across nodes of the power time-courses for TUN reveals significant $1/f$ scaling and infra-slow fluctuations (Suppl. Fig.~\ref{fig:fig_S4}B) -- mirroring the \qty{0.01}{\Hz} components observed in the network-level PSD (Fig.~\ref{fig:fig_2}K).
This pattern of fluctuations in the TUN configuration also aligns with empirical EEG recordings: the PSD of the alpha power time-course in real EEG data shows a comparable $1/f$ scaling (Suppl. Fig.~\ref{fig:fig_S4}B), and the distribution of $\mathrm{CV}_{\mathrm{power}}$ across channels is more consistent with the TUN configuration than with DEF (Suppl. Fig.~\ref{fig:fig_S4}C).

The dual manifestation of infra-slow rhythms across both nodal power time-courses and network-level activity suggests a scale-free spatio-temporal organization, where infra-slow fluctuations coherently modulate the dominant alpha-band dynamics in TUN.
In stark contrast, DEF exhibits a flat PSD profile for power time-courses, consistent with uncorrelated (white-noise-like) dynamics. Together with the $1/f$ scaling (Fig.~\ref{fig:fig_2}K), these findings suggest that TUN may operate near a critical regime. To test this hypothesis, we perform Detrended Fluctuation Analysis (DFA) (see Sec.~\ref{subsubsec:mat_meth:metrics:time_freq_analysis} for details) on the \qtyrange[range-phrase=--]{8}{12}{\Hz} (\qtyrange[range-phrase=--]{18}{22}{\Hz} for DEF) amplitude envelope. DFA exponents in TUN (Fig.~\ref{fig:fig_4}I) span a broad range (exponent $\approx$ \numrange[range-phrase=--]{0.5}{1.2}; IQR = \numrange[range-phrase=--]{0.6}{0.93}), encompassing uncorrelated signals (\num{\approx 0.5}), weak to strong long-range temporal correlations ($\approx$ \numrange[range-phrase=--]{0.6}{1.0}), and persistent trends (> \num{1.0}). This heterogeneity suggests distinct dynamical sub-populations with a specific spatial distribution (Fig.~\ref{fig:fig_4}K): approximately \qty{30}{\percent} of nodes fall within the range \numrange[range-phrase=--]{0.8}{1.0}, often associated with near-critical or critical dynamics and also in line with empirical values observed in EEG data (Suppl. Fig.~\ref{fig:fig_S4}D), while others exhibit either noise-like (\num{\approx 0.5}; \qty{\sim 16}{\percent} of nodes) or more persistent (\num{> 1.0}; \qty{\sim 17}{\percent} of nodes) behavior. In contrast, DEF consistently shows noise-dominated dynamics (exponent $\sim$ \numrange[range-phrase=--]{0.5}{0.6}) across all nodes (Figs.~\ref{fig:fig_4}J and~\ref{fig:fig_4}L).

\subsection{Emergent complex and asymmetric interactions in the tuned configuration}
\label{subsec:results:emergent_interactions}

Up to this point, we have analyzed the dynamics of individual nodes, without examining their mutual interactions. To this specific aim, by leveraging cross-correlation functions one can derive two key metrics: the functional connectivity ($\mathbb{FC}$) and the functional lag ($\mathbb{FL}$) between each pair of nodes (see Sec.~\ref{subsubsec:mat_meth:metrics:corr_funcs} for details). This procedure is illustrated in Fig.~\ref{fig:fig_5}A, where, for a given pair of nodes $i$ and $j$, the cross-correlation between their time series is computed at different time lags.

\begin{figure}[!b]
    \centering
    \includegraphics[width=\linewidth]{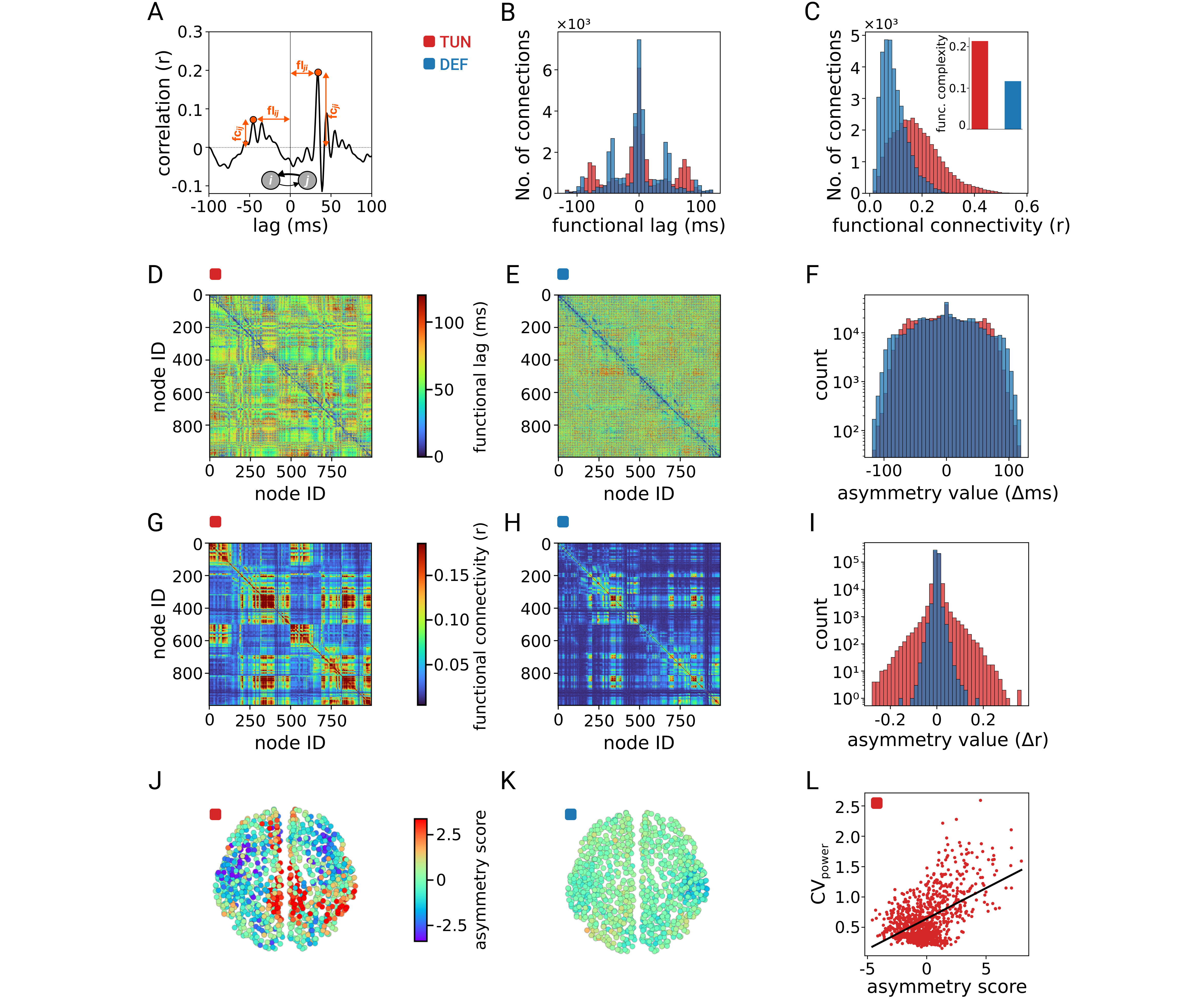}
    \caption{
        \textbf{Functional connectivity, functional lags, and asymmetry in TUN and DEF configurations.}\\
        \textbf{A)} Illustration of the cross-correlation function applied to investigate pairwise interactions between nodes. The correlation between the time series of nodes~$i$ and~$j$ is computed at various time lags. Negative lags indicate a directional relation from~$i$ to~$j$, while positive lags indicate the other way round. The functional connectivity ($\mathbb{FC}$) is defined as the height of the maximum correlation peak, with the corresponding lag being the functional lag ($\mathbb{FL}$).
        \textbf{B)} Distributions of functional lag values across all (ordered) node pairs for
     }
    \label{fig:fig_5}
\end{figure}

\begin{figure}[!t]
    \centering
    \ContinuedFloat
    \caption*{
        the two configurations. \textbf{C)} Distributions of functional connectivity values across all (ordered) node pairs. TUN exhibits a broader $\mathbb{FC}$ distribution compared to DEF, indicating stronger and richer interactions. The inset shows the value of functional complexity.
        \textbf{D-E)} $\mathbb{FL}$ matrices (absolute values for the entries) for TUN and DEF configurations.
        \textbf{F)} Distributions of asymmetry values in $\mathbb{FL}$ matrices, derived from $\mathbb{FL}_{\mathrm{asym}}$.
        \textbf{G-H)} $\mathbb{FC}$ matrices for TUN and DEF configurations.
        \textbf{I)} Distributions of asymmetry values in $\mathbb{FC}$ matrices, derived from $\mathbb{FC}_{\mathrm{asym}}$.
        \textbf{J-K)} Nodal asymmetry scores $\mathbb{A}$ derived from $\mathbb{FC}$ for TUN and DEF configurations.
        \textbf{L)} Linear regression ($r=0.51$, $p=10^{-66}$) between nodal asymmetry scores $\mathbb{A}$ and $\mathrm{CV}_{\mathrm{power}}$ in TUN. Each point represents a brain node.
    }
\end{figure}

The distributions of $\mathbb{FC}$ and $\mathbb{FL}$ values across all node pairs are reported in Figs.~\ref{fig:fig_5}B and~\ref{fig:fig_5}C. Coherently with what seen so far, TUN exhibits higher and more dispersed $\mathbb{FC}$ values than DEF, reflecting stronger and richer interactions.

It has been suggested \citep{zamoralopez2016functional} that the richness of $\mathbb{FC}$ values can serve as an index of functional complexity (see Sec.~\ref{subsubsec:mat_meth:metrics:complexity_metrics}), where the coexistence of high and low $\mathbb{FC}$ values supports a balance between functional integration (high $\mathbb{FC}$ values) and functional segregation (low $\mathbb{FC}$ values). We quantify this complexity during spontaneous activity, confirming that TUN exhibits higher functional complexity compared to DEF (inset of Fig.~\ref{fig:fig_5}C).

The Functional Lag ($\mathbb{FL}$) (absolute values) and the Functional Connectivity($\mathbb{FC}$) matrices are displayed in Figs.~\ref{fig:fig_5}D-E and Figs.~\ref{fig:fig_5}G-H, respectively, alongside with the corresponding asymmetry values in Figs.~\ref{fig:fig_5}F and~\ref{fig:fig_5}I, derived from $\mathbb{FC}_{\mathrm{asym}}$ (see Suppl. Fig.~\ref{fig:fig_S5}) and $\mathbb{FL}_{\mathrm{asym}}$ (see Sec.~\ref{subsubsec:mat_meth:metrics:corr_funcs} for further details). Notably, $\mathbb{FC}$ shows a greater asymmetry in TUN, compared to DEF configuration (IQR = \num{0.01} for TUN; IQR = \num{0.006} for DEF), highlighting a prominent emergence of directional interactions in the tuned configuration. This reflects an emergent network phenomenon, because TUN differs from DEF in single-node equations parameters, global coupling and global conduction velocity, without modifying the relative connection strengths and the time delays of the underlying connectome.

In Figs.~\ref{fig:fig_5}J and~\ref{fig:fig_5}K, we then quantify the asymmetry score $\mathbb{A}$ for each brain node in TUN and DEF, respectively: starting from $\mathbb{FC}_{\mathrm{asym}}$, we marginalize its values along each row to obtain a single asymmetry score per node (see Sec.~\ref{subsubsec:mat_meth:metrics:corr_funcs} for details). Positive values indicate nodes that predominantly exert influence on others (``leading'' or ``parent'' nodes), while negative values denote nodes that are primarily influenced by others (``following'' or ``child'' nodes). First, we observe differences between the left and the right hemisphere, with the right hemisphere expressing more ``leading'' nodes. Then, mapping the asymmetry scores of the nodes to brain regions and sorting them (see Suppl. Fig.~\ref{fig:fig_S6}), we find that the posterior cingulate cortex (PC) has the highest positive asymmetry value in both the left and right hemispheres for the TUN configuration. This result aligns with experimental literature showing that the PC region is a primary driver of the spontaneous brain activity \citep{coito2018directed}. Suppl. Fig.~\ref{fig:fig_S6} further highlights the differences between the left and the right hemisphere, already illustrated in Fig.~\ref{fig:fig_5}J. Once again, DEF shows a more homogeneous behavior, with very little asymmetry values across the whole brain.

In Fig.~\ref{fig:fig_5}L, eventually, we examine the relationship between the asymmetry score and power fluctuations, measured by $\mathrm{CV}_{\mathrm{power}}$. Notably, we find a significant positive correlation ($r=0.51$, $p=10^{-66}$), indicating that nodes with greater power variability tend to have stronger leading roles for the TUN configuration. In contrast, the same analysis for DEF reveals a non-significant relationship ($r=0.009$, $p=0.77$) (see Suppl. Fig.~\ref{fig:fig_S7}).

\subsection{Evoked responses reveal higher spatiotemporal complexity in the tuned configuration}
\label{subsec:results:evoked_responses}

\begin{figure}[!t]
    \centering
    \includegraphics[width=\linewidth]{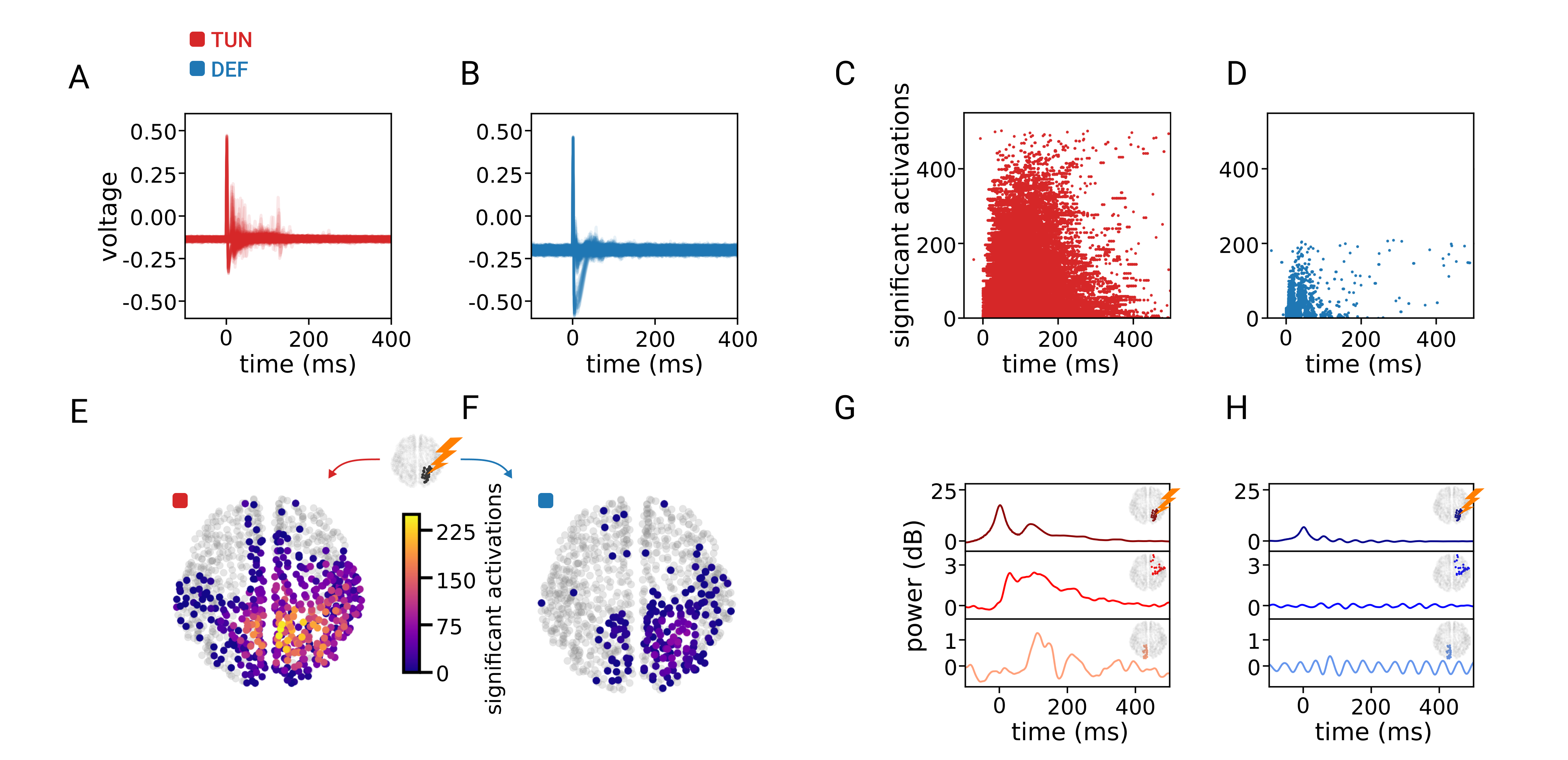}
    \caption{
        \textbf{Stimulus-evoked activity in TUN and DEF configurations.}\\
        \textbf{A-B)} \num{200} independent \qty{2}{\ms} perturbations (i.e. \num{200} trials) were applied to the right superior parietal nodes (rSP) of the brain model, for both TUN and DEF configurations, to assess the spatio-temporal complexity of the evoked activity. Voltage traces averaged across trials in TUN (A, red) and DEF (B, blue) are here reported for each of the \num{998} nodes.
        \textbf{C-D)} Binary spatio-temporal matrices of significant activations across nodes (y-axis) and time (x-axis), showing more sustained and spatially distributed activity in TUN (C) relative to DEF (D), where activations are sparse and temporally constrained.
        \textbf{E-F)} Brain maps showing the spatial distribution of significant activations (sum over \qtyrange[range-phrase=--]{10}{500}{\ms} post-stimulus) across nodes. The smaller brain on top of the colorbar illustrates the stimulated nodes located on the rSP cortex. In TUN (E), the stimulus spreads across distant cortical areas, while in DEF (F), the activity remains localized around the stimulation site.
        \textbf{G-H)} Post-stimulus broadband power (\unit{\dB}) averaged across three equal-sized node subsets (\num{27} nodes): directly stimulated nodes (top), nearby connected but non-stimulated nodes (middle), and distant connected nodes (bottom). Insets indicate node locations. TUN (G) displays higher power and greater temporal variability than DEF (H), where responses are weaker and more stereotyped.
     }
    \label{fig:fig_6}
\end{figure}

To assess the model capacity to sustain complex stimulus-evoked dynamics, we apply \num{200} independent brief (\qty{2}{\ms}) perturbations to the right superior parietal (rSP) nodes in both TUN and DEF configurations. The resulting evoked activity reveals striking differences between the two: in the TUN configuration, the average post-stimulus voltage exhibits richer and less stereotyped dynamics (Fig.~\ref{fig:fig_6}A), compared to the more uniform and rapidly decaying responses observed in the DEF configuration (Fig.~\ref{fig:fig_6}B).

To characterize the spatio-temporal profile of the evoked activity, we derive a binary spatio-temporal matrix of significant activations (see Sec.~\ref{subsubsec:mat_meth:metrics:complexity_metrics} for details). In TUN, the evoked activity is sustained over a broader temporal and spatial window (Fig.~\ref{fig:fig_6}C), whereas in DEF it remains spatio-temporally confined to the immediate post-stimulus period and to the surroundings of the perturbed region (Fig.~\ref{fig:fig_6}D). The prolonged perturbation effects in the TUN configuration cannot be explained solely by the reduction of the timescale parameter: while $t_{\mathrm{scale}}$ decreases by less than half (from \num{1.0} to \num{0.6}) compared to DEF, perturbations in TUN persist nearly \num{2.5} times longer. The broader spatial extent of the TUN response becomes especially evident when the number of significant activations is reported onto the brain maps, suggesting that in TUN the stimulus propagates far from the stimulation site, recruiting distant regions of the network (Fig.~\ref{fig:fig_6}E). In contrast, the DEF response remains tightly localized around the stimulation site (Fig.~\ref{fig:fig_6}F).

To further explore the spread and heterogeneity of the evoked activity, we compute the post-stimulus broadband power (see Sec.~\ref{subsubsec:mat_meth:metrics:time_freq_analysis}) in three equal-sized node subsets: the directly stimulated nodes; a set of spatially nearby but non-stimulated nodes within the same hemisphere; and the most distant nodes connected to the stimulation site, as determined by structural connectivity (i.e., tract lengths). In all three subsets, the TUN configuration exhibits higher power overall, as well as greater variability in its temporal dynamics compared to the DEF configuration (Figs.~\ref{fig:fig_6}G vs~\ref{fig:fig_6}H).

Finally, we quantify the overall complexity of the evoked spatiotemporal patterns using the perturbational complexity index (PCI; see Sec.~\ref{subsubsec:mat_meth:metrics:complexity_metrics}). TUN reaches $\text{PCI} = 0.54$, a substantially higher value with respect to $\text{PCI} = 0.39$ observed for DEF, confirming its ability to sustain richer, more spatially distributed and temporally diverse patterns of activity in response to brief perturbations.
Notwithstanding an absolute comparison with PCI values from literature is not reliable, due to strong dependencies on the experimental protocols for the EEG recording and on the inverse model employed, still the relative difference between the two PCI values found here are consistent with experimentally ones observed in healthy control individuals and subjects with reduced levels of consciousness (either responsive or not) \citep[see, e.g.,][]{casarotto2016stratification}.

\section{Discussion}
\label{sec:discussion}

This study grounds on two fundamental questions: \textit{i)} How can a brain network model be tuned to simultaneously reproduce key characteristics of both spontaneous and externally evoked neural activity? \textit{ii)} How can we systematically evaluate the model consistency with expectations from biology and ultimately with empirical data? This work contributes to answering these questions by defining physiologically relevant metrics that are useful in guiding the exploration and the initial tuning of model parameters, thus setting up a convenient starting point for the calibration process. The approach is iterative, where discrepancies between features exhibited by the model and biological datasets provide valuable feedback that helps refine the model parameters and improve its accuracy in reproducing the neural dynamics. These metrics could be applied equally to synthetic and experimentally obtained data to yield directly comparable results, thereby strengthening the robustness of the model and exploring the boundaries of its applicability. Additionally, to enhance the reproducibility and facilitate cross-study comparisons, these metrics should be accessible to the broader neuroscience community and generalizable to accommodate diverse experimental contexts.

To achieve this, we leverage TVB, a versatile software framework that enables the simulation of large-scale brain network models. In this study, we use TVB to simulate whole-brain dynamics in both spontaneous and evoked conditions, exploring how different parameter configurations of the same model shape the emergent neural activity. However, extracting meaningful insights from these simulations and quantitatively comparing them to empirical findings requires dedicated tools for their processing and analysis, and then back to the model for its calibration. Here, we use Cobrawap to process the TVB simulation output for tuning model parameters, aiming at matching expectations from empirical neurophysiological patterns. This approach enables meaningful comparisons between the default TVB configuration of the model and its tuned counterpart. Cobrawap capabilities for monitoring, analyzing, and visualizing dynamic variables are instrumental in ensuring that the eventually selected network model configuration captures both spontaneous and evoked brain activity. Finally, this approach proves to be essential not only for guiding the tuning process, but also for revealing significant differences in the inner dynamics of the two model configurations, thus allowing for a deeper understanding of the model itself.
Notably, both TVB and Cobrawap are delivered to the EBRAINS community as \textit{tools} of the EBRAINS Software Distribution (ESD) \citep{ebrains_ESD}. Therefore, representing a showcase of their integration, this work is also a proof of concept towards the construction of integrated workflows in EBRAINS.

In the following, we summarize the main outcomes obtained in this work, and we drive to conclusions complementing their interpretation with findings and results from literature.
First, tuning is performed such that an alpha-band peak can be observed, consistent with empirical evidence highlighting the prominence of alpha rhythms during resting-state brain activity \citep{klimesch1999eeg}. This effect results from tuning the global timescale of the model dynamics, while at variance alpha oscillations are absent in the default model.

Second, the tuning also significantly influences both the spectral and spatial organization of brain activity. The analysis of the temporal fluctuation of transition events from the baseline to a higher activity regime and power spectral content reveals a clear spatial organization in the tuned configuration, characterized by marked heterogeneity -- a feature consistent with more realistic brain dynamics \citep{zhang2025modeling}. In contrast, the default configuration exhibits more homogeneous and stereotyped dynamics, with reduced spatio-temporal variability in both events and power. Notably, these differences stem from the tuning of the model equation parameters, the global coupling strength, and the conduction velocity (as both configurations are based on the same underlying connectome, i.e. structural connectivity: relative weights of synaptic connections are not changed).

Mapping the auto-correlation peak latencies onto the brain regions further highlights the non-trivial spatial organization of the tuned model, where faster periodicities are observed in frontal regions, whereas slower oscillations are present in posterior regions. This spatio-temporal gradient is consistent with experimental observations \citep{rosanova2009natural, frauscher2018atlas, capilla2022natural}. Interestingly, also centro-parietal regions exhibit auto-correlation peak latencies shorter than the median periodicity of other cortical regions, thus forming a distinct cluster of cortical nodes. These populations also display lower average event rates, indicating fewer periods of neural mass synchronization. This pattern corresponds to more differentiated and fluctuating dynamics, as reflected by the increased variability in their power time-course.

Furthermore, our study examines the communication between nodes using cross-cor\-re\-la\-tion functions. We extract two relevant metrics to analyze nodal interactions: functional connectivity ($\mathbb{FC}$) and functional lag ($\mathbb{FL}$). Our analysis reveals that the tuned configuration of the model exhibits a wider distribution of functional connectivity ($\mathbb{FC}$) values compared to the default one, suggesting a greater diversity of interactions in the former. This reflects a characteristic of complex brain dynamics, that maintains a balance between integration and segregation across the network \citep{zhao2010complexity, zamoralopez2016functional}.

The emergence of a greater $\mathbb{FC}$ asymmetry in the tuned configuration is another key finding. We find more directional interactions in the tuned model with respect to the default behavior, underscoring the effect of parameter tuning in promoting asymmetric network dynamics. In turn, this could indicate the presence of emergent network motifs that are crucial for sustaining complex neuronal patterns \citep{coito2018directed, monma2025directional}. Notably, centro-parietal regions with higher asymmetry scores also exhibit greater power fluctuations, as reflected by a significant correlation between asymmetry scores and the $\mathrm{CV}_{\mathrm{power}}$. This finding suggests that nodes with stronger leading roles in the network exhibit more variable neural activity, potentially enhancing their ability to coordinate information propagation, thus acting as \textit{hubs}. Such variability may drive neuronal cascades -- transient bursts of activity that could shape communication across large-scale brain networks \citep{rabuffo2021neuronal, fousek2024symmetry}. Notably, this asymmetry is particularly pronounced in the posterior cingulate cortex, consistent with previous experimental evidence \citep{coito2018directed}.

Also, we observe the emergence of infra-slow rhythms (\qty{\sim 0.01}{\Hz}) in the tuned configuration, well-documented in experiments \citep{palva2012infraslow, gutierrezbarragan2019infraslow}. A particularly interesting question arising from these findings is whether the observed spatial heterogeneity in terms of temporal variability influences these infra-slow fluctuations. Specifically, it remains unclear whether centro-parietal regions, already identified as potential hubs because of their high temporal variability and stronger leading role, act as ``orchestrators'' of these network infra-slow rhythms, or whether they act as ``gates'' that facilitate the propagation of fronto-posterior (or postero-frontal) waves across the network. This question, which we intend to explore in future research, offers a promising direction for investigating how specific regions interact with infra-slow dynamics and contribute to the reconfiguration of brain states within large-scale networks \citep{rabuffo2021neuronal}.

Crucially, the results on spontaneous activity are also accompanied by evidence that properly tuning the model parameters can generate complex, stimulus-evoked dynamics. In line with empirical studies of brain activity \citep{casali2013theoretically, sarasso2014quantifying}, we find that the model is able to sustain complex spatio-temporal patterns of activation in response to external perturbations, quantified with the PCI. Notably, in this work we reduce the stimulus duration from \qty{5}{\ms} to \qty{2}{\ms} -- compared to the protocol used in~\citet{gaglioti2024investigating} -- to better match experimental conditions, including both intra- and extra-cranial stimulations \citep{comolatti2025transcranial}. Despite the shorter stimulus, the model still exhibits a realistic complex response, further corroborating its ability to capture essential features of neural dynamics beyond spontaneous fluctuations.

A theoretical framework able to integrate our results on spontaneous and evoked dynamics is provided by criticality -- a dynamical regime emerging near phase transitions, where systems exhibit a unique balance between ordered and disordered activity \citep{beggs2003neuronal, chialvo2010emergent, obyrne2022how}. In this regime, scale-free properties naturally arise, collectively maximizing information capacity \citep{shew2012functional}: $1/f$ spectral scaling \citep{he2010temporal}, long-range temporal correlations \citep{linkenkaerhansen2001longrange, dallaporta2019modeling}, infra-slow fluctuations \citep{palva2012infraslow, palva2018roles}, and heightened sensitivity to external perturbations \citep{obyrne2022how}. In the tuned configuration, such signatures emerge naturally in both spontaneous and evoked activity. In particular, the heightened sensitivity of TUN to brief external stimuli -- manifested as prolonged and complex responses, with longer relaxation timescales -- resembles the amplified perturbation effects typical of the \textit{critical slowing down} phenomenon \citep{meisel2015critical}. In addition, the richness of spatio-temporal patterns measured with PCI suggests a more efficient signal propagation through the network. Together with empirical links between PCI and criticality \citep{maschke2024critical}, these results seem to indicate that parameter tuning shifts the model toward a critical-like regime, closer to the dynamical richness of biological systems.

\subsection{Limitations and perspective}
\label{subsec:discussion:limits_persp}

Our study showcases how to tune models to capture macroscopic features of brain dynamics, however it is important to acknowledge some limitations. First, we focus on reproducing large-scale characteristics of neural activity as resulting from the corpus of experimental evidence. As a proof of concept of the comparability with experimental data, we propose a close comparison between simulated and empirical features in Suppl. Fig.~\ref{fig:fig_S8}, which also demonstrates the generalizability of our approach to other connectomes (specifically, a \num{74}-node connectome).
Further steps toward an accurate calibration between TVB outputs and extra-cortical recordings (like hd-EEG) would require refined forward models \citep{gramfort2013meg}.
Nevertheless, our approach lays the foundation for such quantitative calibration by adopting metrics that have proved effective in assessing experimental features in the Cobrawap framework \citep{debonis2019analysis, celotto2020analysis, capone2023simulations, gutzen2024modular}.


Secondly, the Larter-Breakspear model has been designed for reproducing the dynamics of cortical columns and does not include detailed contributions from subcortical inputs, which conversely are known to play a fundamental role in orchestrating brain rhythms and propagating neural activity. Some recent works \citep[e.g.,][]{lorenzi2025region} tackle this task with TVB-based simulations, offering indications for incorporating some advances in the Larter-Breakspear framework.
Nevertheless, even if the TVB simulation engine leverages simplified representations of cortical dynamics as neural mass models that describe synchronization processes at a population level, this abstraction still enables the study of several physiological and pathological conditions.
Indeed, neural mechanisms relevant for brain diseases can be modeled even just by acting on easily interpretable global parameters, as the conduction speed or the global coupling \citep{deco2013resting, kringelbach2015rediscovery, falcon2016functional, stefanovski2019linking, aerts2020modeling}, once more evidencing the importance of an accurate tuning.

Additionally, TVB simulations can be the basis for surgical and clinical treatments, like epilepsy: see the experience of EPINOV, the world’s first clinical trial on predictive brain modeling in epilepsy surgery~\citep{jirsa2023personalized}. Again, careful parameter tuning -- including personalization, toward a digital twin approach~\citep{wang2025virtual, martin2026tvbcplusplus} -- is crucial for reliable and effective results.


The Larter–Breakspear model employed here has been extensively used for simulating fMRI BOLD signals \citep{honey2007network, honey2009predicting, alstott2009modeling, roberts2019metastable, endo2020evaluation}. Neural mass activity can be converted into BOLD signals using the nonlinear Balloon–Windkessel hemodynamic model, which translates neuronal dynamics into blood-oxygen-level-dependent responses \citep{friston2000balloon}. This model is readily available as a built-in monitor in TVB platform \citep{sanzleon2013virtual}, and its integration into our modeling framework is straightforward, as demonstrated by \citet{gaglioti2024investigating}. Therefore, extending our current framework to simulate fMRI data is an easily implementable step in future work, enabling direct validation against empirical resting-state fMRI recordings and opening new avenues for clinical applications and brain-computer interface development \citep{sitaram2007fmribci}.

Finally, our analysis primarily focuses on a stationary characterization of the model dynamics, identifying fixed patterns of spontaneous activity. However, the brain does not operate in a steady state -- it exhibits transient dynamics, shifting between different functional configurations over time. Future studies will aim to extend this work by exploring the temporal evolution of network states, classifying these transient states, and ultimately comparing them with empirical data. This will further improve model calibration and help refine our understanding of how large-scale neural networks reconfigure over time.

\subsection{Conclusion}
\label{subsec:discussion:concl}

In summary, this study demonstrates the importance of parameter tuning in capturing the full complexity of spontaneous and evoked brain dynamics. The use of Cobrawap is instrumental in refining the model and ensuring its consistency with empirical expectations, allowing us to reproduce key brain activity features such as alpha-band oscillations, infra-slow rhythms, spatio-temporal heterogeneity in network activity and complex evoked responses.
The independently developed metrics, applicable to any simulation engine, enable a more systematic approach that creates opportunities for iterative automatic data-driven refinement of model parameters through tools like the Learning-to-Learn (L2L) framework \citep{yegenoglu2022exploring}. 
Our findings establish a proof of concept for a structured methodology that bridges computational simulations and experimental data in humans. This paves the way for quantitative calibration and validation of accurate, interpretable, reliable and personalized models of whole-brain dynamics, in both healthy and diseased conditions.

\section*{CRediT authorship contribution statement}


\noindent
\textbf{Gianluca Gaglioti:} Conceptualization, Formal analysis, Investigation, Methodology, Software, Visualization, Writing - Original Draft, Writing - Review \& Editing. \textbf{Alessandra Cardinale:} Formal analysis, Methodology, Software, Visualization, Writing - original draft, Writing - review \& editing. \textbf{Cosimo Lupo:} Conceptualization, Formal analysis, Investigation, Methodology, Software, Supervision, Writing - original draft, Writing - review \& editing. \textbf{Thierry Nieus:} Conceptualization, Formal analysis, Investigation, Methodology, Supervision, Writing - original draft, Writing - review \& editing. \textbf{Federico Marmoreo:} Methodology, Software, Writing - review \& editing. \textbf{Elena Focacci:} Data curation, Methodology, Writing - review \& editing. \textbf{Robin Gutzen:} Methodology, Software, Writing - review \& editing. \textbf{Michael Denker:} Methodology, Software, Writing - review \& editing. \textbf{Andrea Pigorini:}  Conceptualization, Funding acquisition, Project administration, Writing - review \& editing. \textbf{Marcello Massimini:} Conceptualization, Funding acquisition, Project administration, Supervision, Writing - review \& editing. \textbf{Simone Sarasso:} Conceptualization, Funding acquisition, Supervision, Writing - review \& editing. \textbf{Pier Stanislao Paolucci:} Conceptualization, Formal analysis, Funding acquisition, Investigation, Methodology, Project administration, Supervision, Writing - original draft, Writing - review \& editing. \textbf{Giulia De Bonis:} Conceptualization, Formal analysis, Investigation, Methodology, Software, Supervision, Writing - original draft, Writing - review \& editing.

\section*{Data and code availability}

Whole-brain simulations rely on software release 2.7.2 of TheVirtualBrain (TVB) (\url{https://www.thevirtualbrain.org/tvb/zwei/brainsimulator-software}), with parameters as in Suppl. Tab.~\ref{tab:params}. Simulation outputs are then analyzed with dedicated tools; their integration in the next official Cobrawap software release 0.3.0 is currently underway (the latest stable release is always reachable at \url{https://doi.org/10.5281/zenodo.10198748}), though new software features developed for this paper are already available as a development branch at \url{https://github.com/APE-group/cobrawap/tree/devs/TVB}. The data from simulations and the code for reproducing the figures presented in this paper are available upon request.

\section*{Fundings}

This work has been co-funded by: the European Next Generation EU through Italian MUR grants CUP B51E22000150006 (EBRAINS-Italy IR00011 PNRR Project) (CL, FM, PSP, GDB, AP, MM); the European Union’s Horizon Europe Programme under the Specific Grant Agreement No. 101147319 (EBRAINS 2.0 Project) (MD, AP); The European Research Council (ERC-2022-SYG-101071900-NEMESIS) (MM); The Tiny Blue Dot Foundation (MM); the Ministero dell’Universit\`a e della Ricerca PRIN 2022 20228ARNXS ``DARKSCANNER'' (SS); the Italian Ministry of Foreign Affairs and International Cooperation, MultiScale Brain Function (MSBFIINE) India-Italy Network of Excellence (DST-MAE) IN22NOE02 (TN); Progetto Di Ricerca Di Rilevante Interesse Nazionale (PRIN) P2022FMK77 (AP). AC is a PhD student enrolled in the National PhD in Artificial Intelligence, XXXVIII cycle BIS, course on Health and life sciences, organized by Universit\`a Campus Bio-Medico di Roma.


\bibliographystyle{elsarticle-harv}
\bibliography{refs_v2}

\clearpage

\appendix

\pagenumbering{gobble} 
\setcounter{figure}{0}
\setcounter{table}{0}
\renewcommand{\thefigure}{A.\arabic{figure}}
\renewcommand{\thetable}{A.\arabic{table}}
\renewcommand{\theequation}{A.\arabic{equation}}
\setcounter{affn}{0}
\resetTitleCounters

\makeatletter
\let\@title\@empty
\makeatother

\title{Emergent complexity and rhythms in evoked and spontaneous dynamics of human whole-brain models after tuning through analysis tools -- Supplementary Material}

\makeatletter
\def\ps@pprintTitle{%
     \let\@oddhead\@empty
     \let\@evenhead\@empty
     \def\@oddfoot{\footnotesize\itshape\hfill}%
     \let\@evenfoot\@oddfoot}
\makeatother

\maketitle

\section*{Supplementary Methods}
\label{sec:suppl_methods}

\subsection*{\textbf{EEG preprocessing, 4 subjects, 62 channels}}

We recorded resting-state electroencephalography (EEG) from \num{4} healthy subjects with eyes closed. Each recording session lasted between \num{5} and \num{10} minutes. EEG measurements were acquired using a \num{64}-channel EEG amplifier (Brain Products GmbH) sampled at \qty{5000}{\Hz} with hardware filtering at \qty{1000}{\Hz}, and electrode impedance was maintained below \qty{10}{\kilo\Omega}. Reference and ground electrodes were placed on the forehead above the frontal sinus.

Resting-state EEG data were preprocessed using the MNE-Python software package \citep{gramfort2013meg}. EEG recordings were first detrended to remove low-frequency drifts. A band-pass filter was applied with a high-pass cutoff at \qty{0.5}{\Hz} and a low-pass cutoff at \qty{60}{\Hz}. Additionally, a \qty{50}{\Hz} notch filter was applied to suppress line noise. Prior to independent component analysis (ICA), time segments containing prominent artifacts, such as those associated with muscle activity or other non-neural sources, were manually removed to improve ICA performance. The ICA decomposition was then applied to identify and remove remaining stereotypical artifacts, such as eye blinks and other physiological sources.

\subsection*{\textbf{Frequency and time-frequency analysis of EEG signals}}

Frequency and time-frequency analyses of the EEG were performed using the MNE-Python package. The power spectral density (PSD) was computed using Welch's method with a Hamming window, \qty{4}{\s} segments and \qty{75}{\percent} (i.e., \qty{3}{\s}) overlap. The PSD was averaged across channels, and both the peak frequency and spectral slope were computed to characterize the aperiodic $1/f$ trend of the signal. The spectral slope was calculated in the \qtyrange[range-phrase=--]{0.7}{7}{\Hz} range using the method described by \citet{colombo2019spectral}, adapted for Python implementation.

Time-frequency analysis was performed by extracting the envelope and instantaneous power time course in the alpha band (\qtyrange[range-phrase=--]{8}{12}{\Hz}) using Morlet wavelet transforms implemented in MNE-Python. To quantify the temporal dynamics of these signals, we computed the coefficient of variation ($\mathrm{CV}_{\mathrm{power}}$) of the power time course across time for each channel and applied detrended fluctuation analysis (DFA) to the envelope signal, as detailed in Sec.~\ref{subsubsec:mat_meth:metrics:time_freq_analysis}.

\subsection*{\textbf{Simulation and analysis using a smaller 74 parcellation connectome}}

As a proof-of-principle demonstration of the generalizability of our framework and its direct applicability to experimental data, we performed additional simulations using a reduced-scale connectome. This supplementary analysis serves two purposes: first, to validate that the parameter tuning approach remains effective when applied to a different parcellation scheme; second, to establish a more quantitative benchmark by directly comparing simulated and empirical neurophysiological features.

We employed a publicly available human connectome provided within the TVB platform. This connectivity matrix has been utilized in prior computational studies \citep{kunze2016transcranial, gaglioti2024investigating} and comprises \num{74} cortical regions, with \num{37} areas per hemisphere. The anatomical parcellation represents a hybrid construction that integrates diffusion spectrum imaging tractography and data derived from the CoCoMac neuroinformatics database \citep{koetter2004online}.

We simulated neural mass dynamics using the Larter-Breakspear model under three distinct parameter configurations. First, we executed the default TVB parameter set to establish a baseline (referred to herein as DEF). Second, we implemented the parameters previously tuned in \citet{gaglioti2024investigating} (referred to herein as TUN1). Third, to further refine the model match to experimental electrophysiology, we conducted a systematic parameter space exploration centered on three key parameters used in the main analysis: global coupling strength ($G$), conduction velocity ($c_v$), and the timescale of the neural mass dynamics ($t_{\mathrm{scale}}$).

The optimal parameter combination was identified by maximizing a composite score that quantifies the correspondence between simulated and empirical spectral characteristics. For each configuration, we computed the average PSD across all regions and extracted its $1/f$ spectral slope using the same estimation procedure described in the previous section on EEG analysis. The score was defined as:
\begin{equation}
    \mathrm{score} \equiv r_{\mathrm{Pearson}}(\mathrm{PSD}_{\mathrm{model}}, \mathrm{PSD}_{\mathrm{EEG}}) - \left(\frac{\mathrm{slope}_{\mathrm{model}} - \mathrm{slope}_{\mathrm{EEG}}}{\mathrm{slope}_{\mathrm{EEG}}}\right)^2
\end{equation}
where $r_{\mathrm{Pearson}}$ denotes the Pearson correlation coefficient between model and EEG PSD, and ``$\mathrm{slope}$'' represents the $1/f$ spectral slope. This metric rewards both spectral shape and spectral slope similarity.
The parameter combination yielding the highest score was $G = 3.6$, $c_v = 4$, and $t_{\mathrm{scale}} = 1$ (referred to herein as TUN2).

Following parameter selection, we computed two metrics of temporal variability for each model configuration: the coefficient of variation of the power time-course at the dominant frequency ($\mathrm{CV}_{\mathrm{power}}$), and detrended fluctuation analysis (DFA) of the envelope. Both metrics were calculated using the identical procedures described for the main analysis. The resulting values were compared directly with the corresponding experimental measurements to evaluate the extent to which TUN2 captures the temporal dynamics observed in real brain activity.

\clearpage

\begin{figure}[p]
    \centering
    \includegraphics[width=\linewidth]{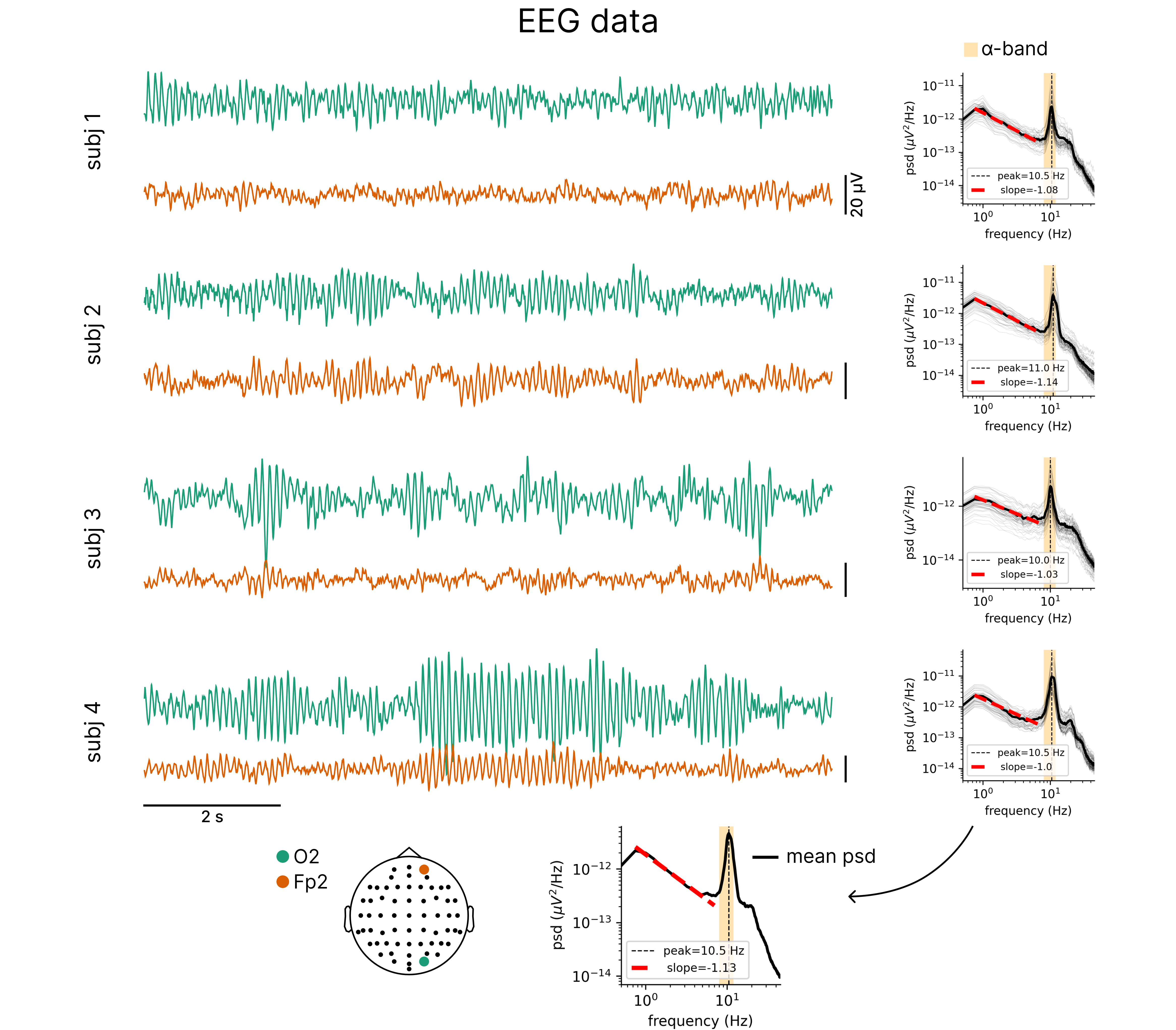}
    \caption{
        \textbf{Representative empirical EEG signals and power spectra during eyes-closed rest}.\\
        Ten-second segments from an occipital (O2, green) and a frontal (Fp2, orange) electrode are shown for each of the four subjects. For each subject, the right panels display the power spectral density (PSD) of all \num{62} channels (thin gray lines) and their average (thick black line). The yellow shading marks the alpha band (\qtyrange[range-phrase=--]{8}{12}{\Hz}), which exhibits a prominent peak in all subjects. The red dashed line shows the spectral slope of the average PSD fitted (as in \citet{colombo2019spectral}) between \num{0.7} and \qty{7}{\Hz}, prior to the alpha peak, to capture the $1/f$ background scaling while avoiding the artificial \qty{0.5}{\Hz} peak introduced by the high-pass filter at \qty{0.5}{\Hz}. The bottom right panel shows the PSD averaged across subjects.
    }
    \label{fig:fig_S1}
\end{figure}

\clearpage

\begin{figure}[p]
    \centering
    \includegraphics[width=\linewidth]{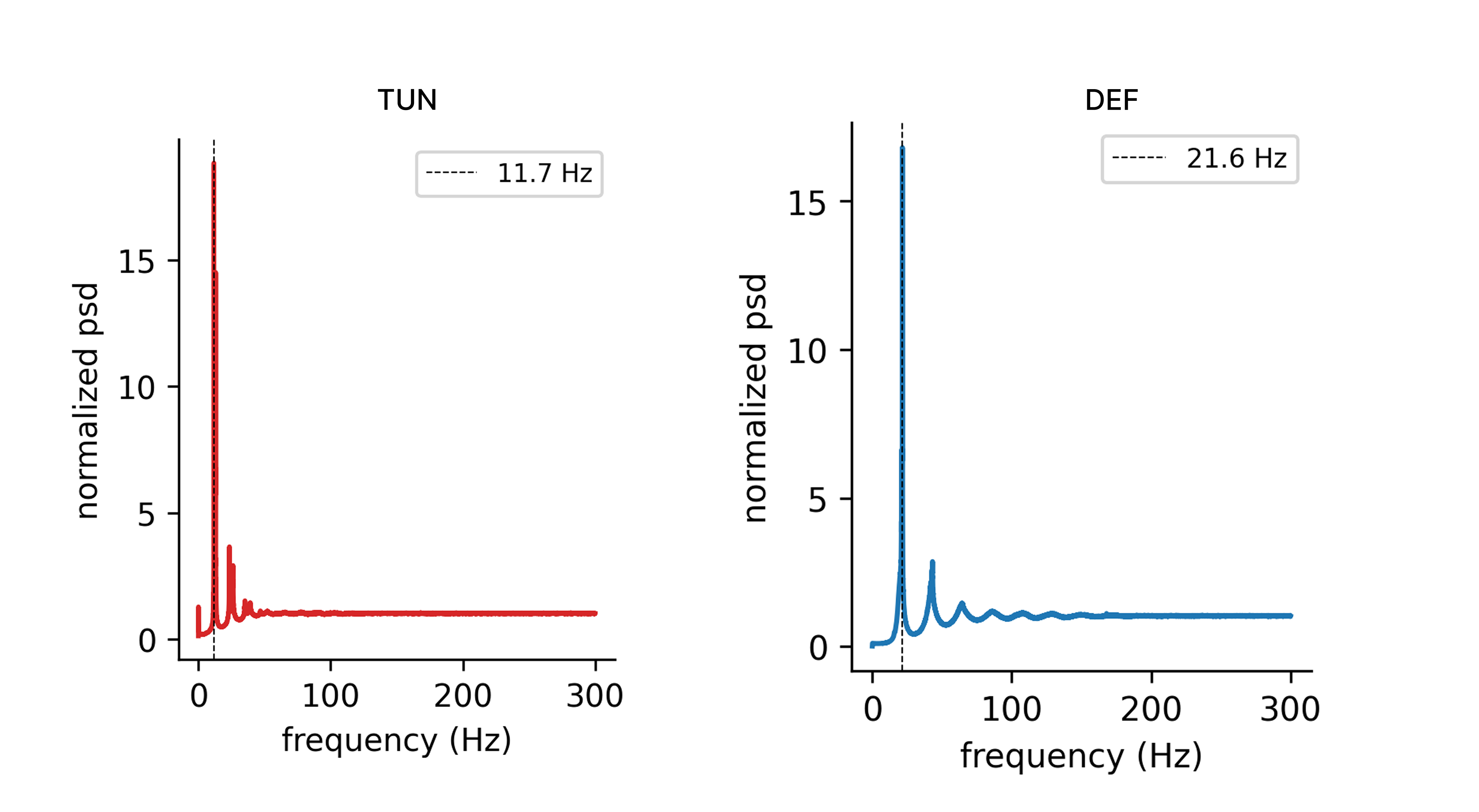}
    \caption{
        \textbf{Power Spectral Density (PSD) of the events in TUN and DEF}.\\
        PSDs are computed on the event time series for each node using the Welch method (with the same parameters as in Fig.~\ref{fig:fig_2}), and then averaged across nodes for TUN and DEF.
    }
    \label{fig:fig_S2}
\end{figure}

\clearpage

\begin{figure}[p]
    \centering
    \includegraphics[width=\linewidth]{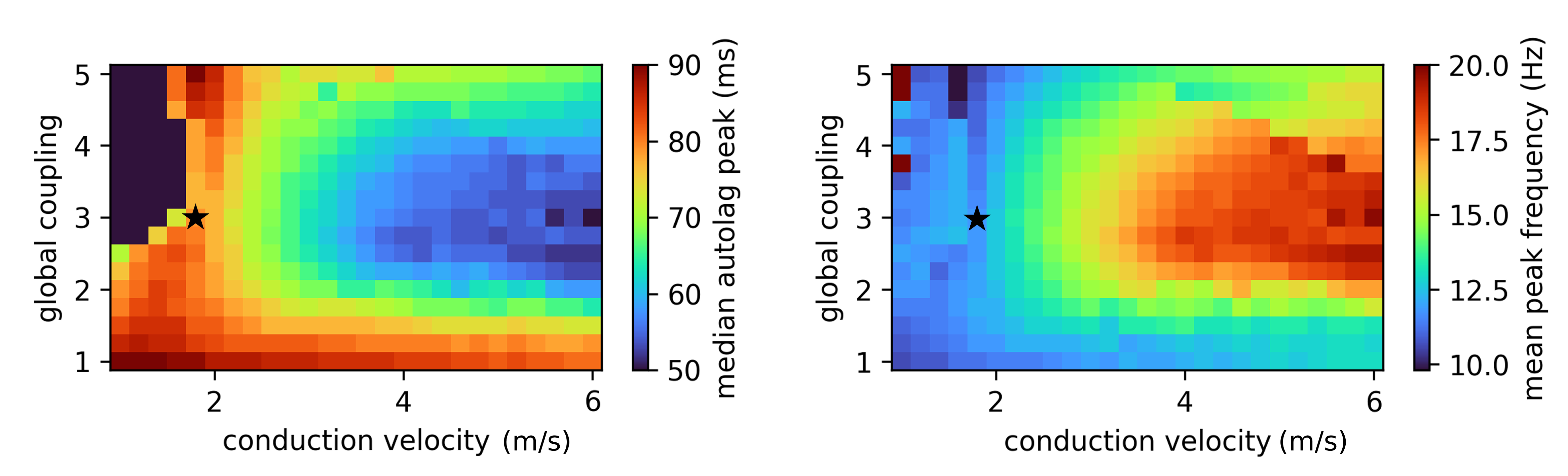}
    \caption{
        \textbf{Global TVB simulation parameters influence the global periodicity of the model}.\\
        We vary the conduction velocity $c_v$ (from \num{1} to \qty{6}{\m/\s}) and the global coupling $G$ (from \num{1} to \num{5}), quantifying the resulting changes in the median auto-correlation peak (second mode), for simulations obtained with TUN configuration for the model. These manipulations induce variations in periodicity: higher conduction velocities are associated with faster periodicities (i.e., shorter median lags) across different global coupling values. The star identifies the actual values eventually used for simulations in this work.
    }
    \label{fig:fig_S3}
\end{figure}

\clearpage

\begin{figure}[p]
    \centering
    \includegraphics[width=0.97\linewidth]{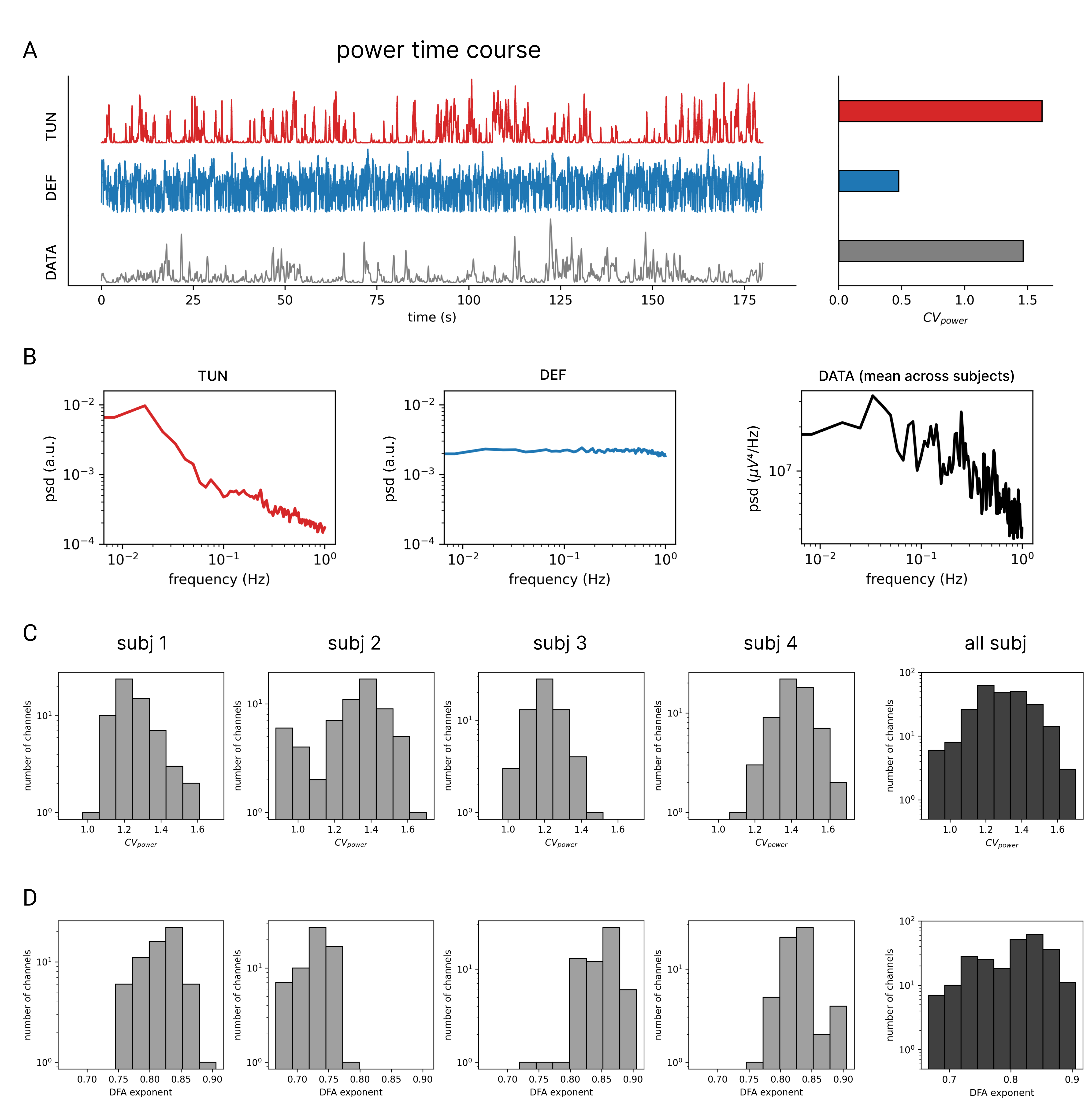}
    \caption{
        \textbf{Signal fluctuations and long-range temporal correlation in EEG data (4 subjects, 62 channels) and model}.\\
        \textbf{A)} The left panel shows the time course of band-limited power, summed over the dominant frequency, in a node of the superior parietal cortex from the TUN configuration (\qtyrange[range-phrase=--]{8}{12}{\Hz}, red), the DEF configuration (\qtyrange[range-phrase=--]{18}{22}{\Hz}, blue), and the occipital electrode O2 from EEG of one subject (\qtyrange[range-phrase=--]{8}{12}{\Hz}, gray). On the right, the coefficient of variation of power $\mathrm{CV}_{\mathrm{power}}$ over time is displayed for each trace, indexing the degree of temporal fluctuation in each condition.
        \textbf{B)} Average power spectral density (PSD) of the power time courses for TUN (red), DEF (blue), and EEG data (black). PSD estimates are computed using \num{120}-\unit{\s} segments with \num{70}-\unit{\s} overlap. For model data, the PSD is averaged across nodes; for EEG, across channels and subjects.
        \textbf{C)} Distribution of $\mathrm{CV}_{\mathrm{power}}$ across channels is shown for each EEG subject (gray), with the aggregated distribution across all channels and subjects shown on the right (black).
        \textbf{D)} Same as C), but for the detrended fluctuation analysis (DFA) exponent, calculated on the alpha envelope as described in Figs.~\ref{fig:fig_4}E-F.
    }
    \label{fig:fig_S4}
\end{figure}

\clearpage

\begin{figure}[p]
    \centering
    \includegraphics[width=\linewidth]{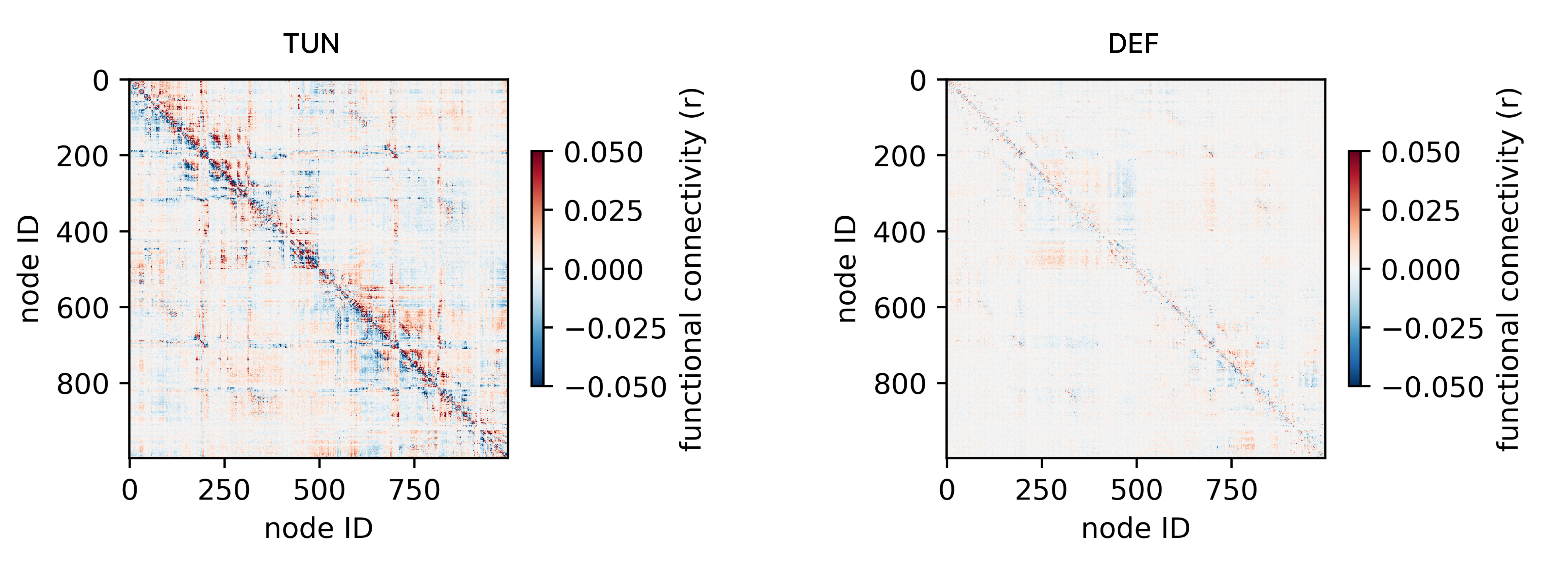}
    \caption{
        \textbf{Functional connectivity asymmetry matrix in TUN and DEF configurations}.\\
        The $\mathbb{FC}_{\mathrm{asym}}$ matrix is quantified as $\mathbb{FC}_{\mathrm{asym}} = \mathbb{FC} - \mathbb{FC}^{\intercal}$, where ${(\cdot)}^{\intercal}$ denotes the matrix transpose. Positive values in $\mathbb{FC}_{\mathrm{asym},ij}$ indicate that node $i$ has a stronger influence on node $j$ than vice versa; negative values indicate the opposite.
    }
    \label{fig:fig_S5}
\end{figure}

\clearpage

\begin{figure}[p]
    \centering
    \includegraphics[width=\linewidth]{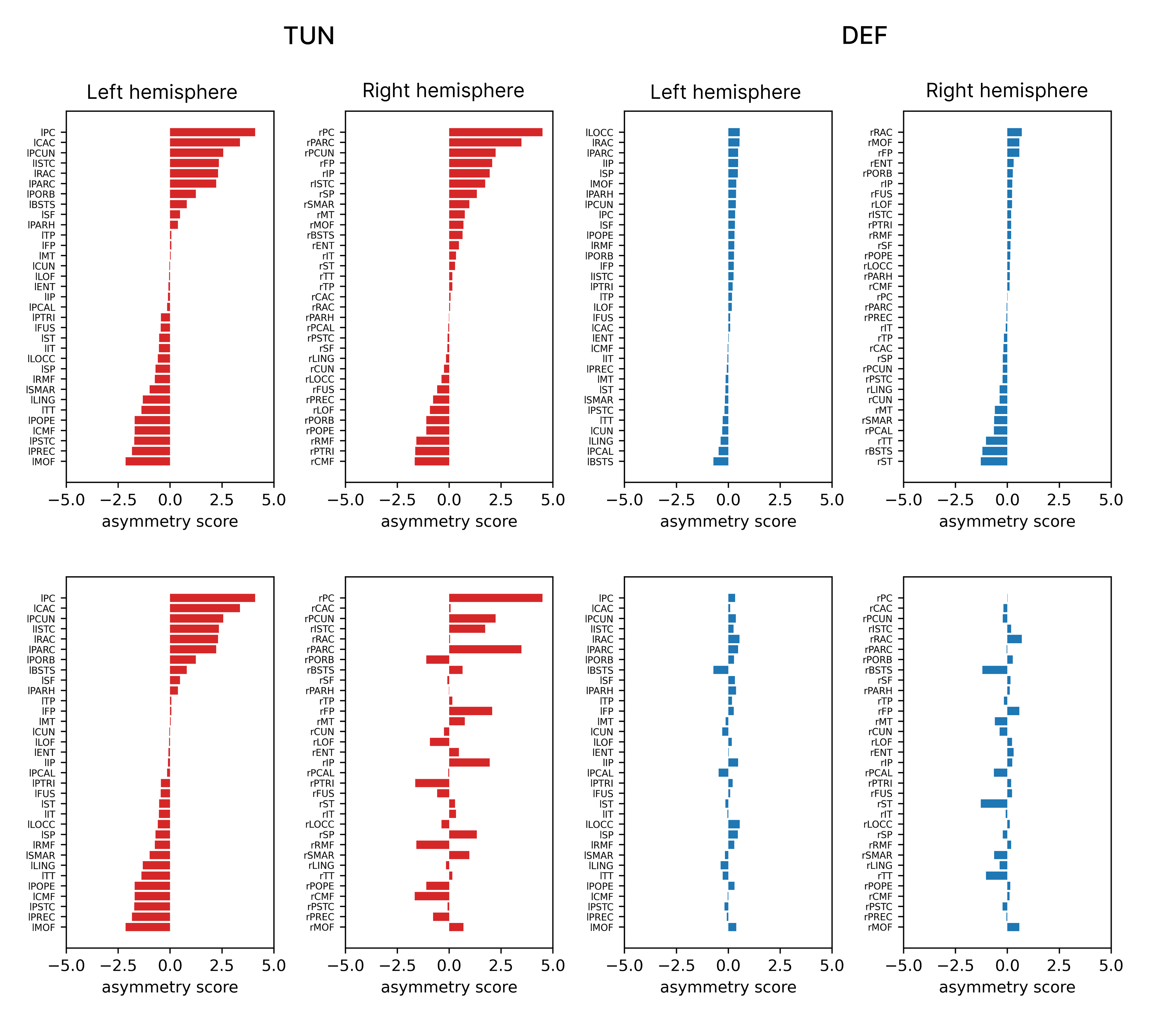}
    \caption{
        \textbf{Average asymmetry score of each cortical region in TUN and DEF}.\\
        Values for listed regions are obtained averaging the asymmetry score across the nodes located in the given region. In top panels, regions are sorted in their respective descending order. In bottom panels, regions are sorted in descending order according to the values of the first panel (left hemisphere of the TUN configuration).
    }
    \label{fig:fig_S6}
\end{figure}

\clearpage

\begin{figure}[p]
    \centering
    \includegraphics[width=0.6\linewidth]{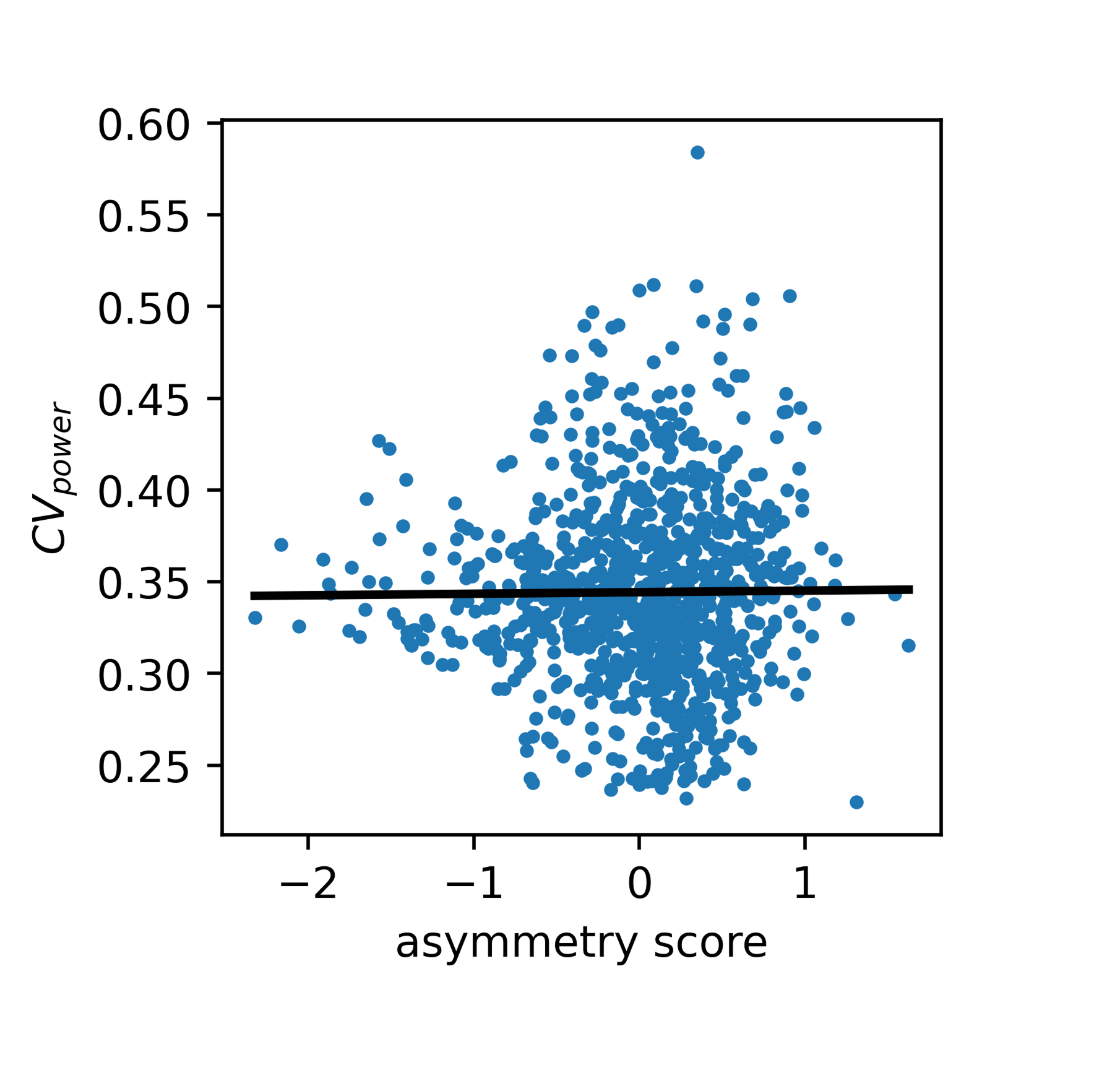}
    \caption{
        \textbf{Relationship between asymmetry score and power fluctuations ($\mathrm{CV}_{\mathrm{power}}$) in DEF}.\\
        Linear regression (solid black line) shows a non-significant correlation ($r=0.009$, $p=0.77$).
    }
    \label{fig:fig_S7}
\end{figure}

\clearpage

\begin{figure}[p]
    \centering
    \includegraphics[width=\linewidth]{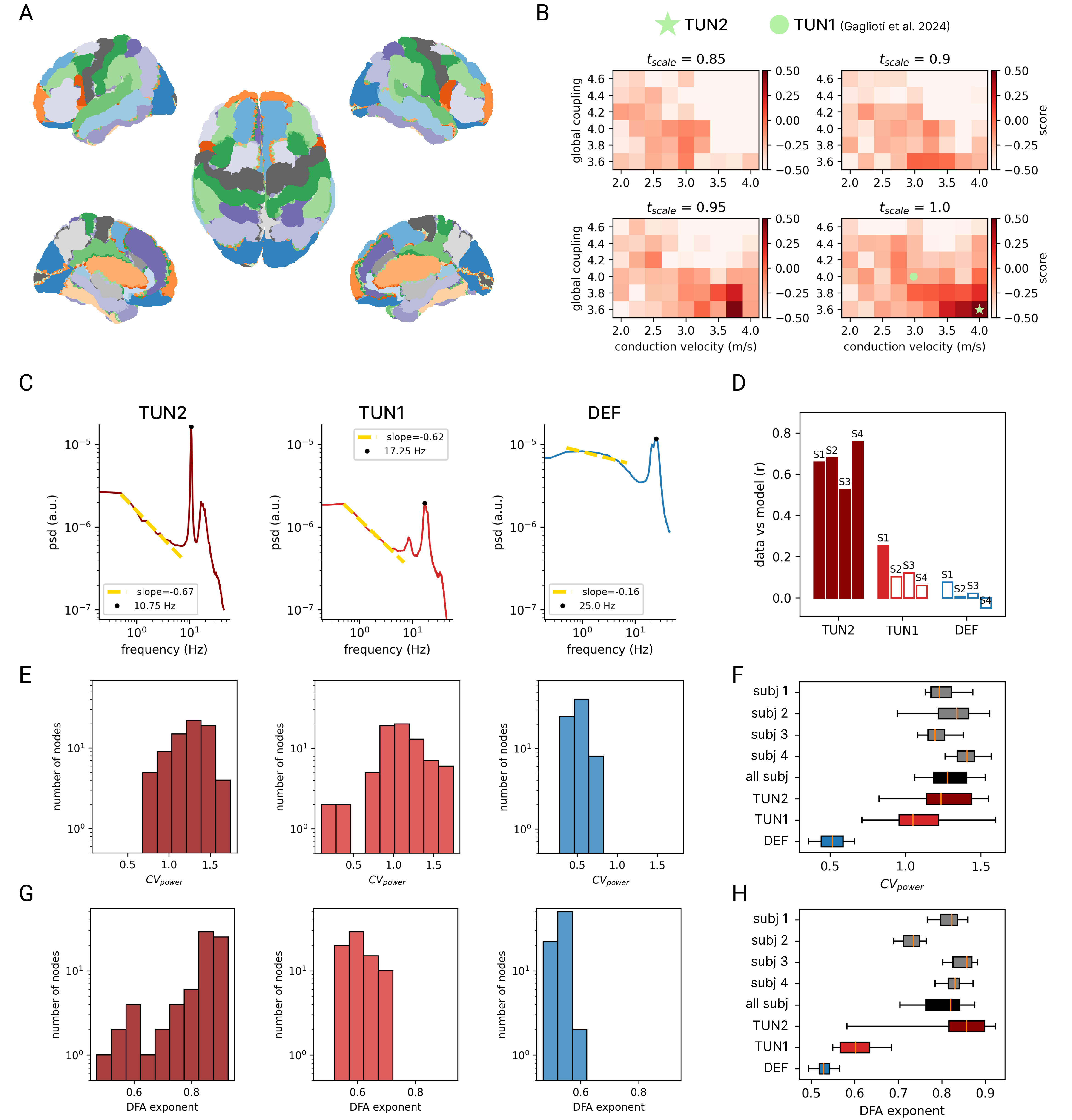}
    \caption{
        \textbf{Tuning a 74-node connectome against empirical EEG data}.\\
        \textbf{A)} Spatial layout of the 74-node parcellation used in the supplementary analysis.
        \textbf{B)} Heatmaps of the parameter space exploration showing the composite score assessing spectral similarity between simulated and empirical PSD. Global coupling strength ($G$, \numrange{3.6}{4.6}, step 0.2), conduction velocity ($c_v$, \numrange{2}{4}, step 0.25), and timescale ($t_{\mathrm{scale}}$, \numrange{0.85}{1.0}, step 0.05) were varied. TUN1 (green dot) refers to the configuration of parameters used in \citet{gaglioti2024investigating}, while TUN2 (green star) indicates the configuration with the maximum score.
        \textbf{C)} Average PSD across regions for the three model configurations: TUN2 (dark red), TUN1 (red), and the default TVB configuration (DEF, blue). The yellow line shows the fitted $1/f$ spectral
    }
    \label{fig:fig_S8}
\end{figure}

\begingroup
\makeatletter
\setlength{\@fptop}{0pt}
\setlength{\@fpbot}{0pt}
\makeatother
\begin{figure}[p]
    \centering
    \ContinuedFloat
    \caption*{
        slope, and black dots mark the spectral peak frequency.
        \textbf{D)} Pearson correlation coefficient ($r$) between model-simulated and empirically measured PSD for individual subjects (S1--S4) across all three model configurations. White-filled bars indicate non-significant correlations ($p > 0.05$).
        \textbf{E)} Distributions of the $\mathrm{CV}_{\mathrm{power}}$ at the dominant oscillation frequency for each configuration. Frequencies are \qtyrange[range-phrase=--]{8}{12}{\Hz} for TUN1, \qtyrange[range-phrase=--]{15}{19}{\Hz} for TUN2, and \qtyrange[range-phrase=--]{23}{27}{\Hz} for DEF, as defined by the spectral peak in panel C.
        \textbf{F)} Boxplots comparing the $\mathrm{CV}_{\mathrm{power}}$ distributions across subjects and models. Gray boxes show individual subject values (S1--S4); black boxes show aggregated data across all subjects. For EEG data, all channels are included; for model data, all regions are included.
        \textbf{G)} Distributions of detrended fluctuation analysis (DFA) exponents computed on the amplitude envelope of the dominant oscillation in each configuration, analogous to panel E.
        \textbf{H)} Boxplots comparing DFA exponents across subjects and models, with the same layout as panel F.
    }
\end{figure}
\clearpage
\endgroup

\begingroup
\setlength{\tabcolsep}{7pt} 
\renewcommand{\arraystretch}{1.15} 
\begin{table}[p]
    \centering
    \caption{
        \textbf{List of cortical regions and number of associated nodes.}\\
        Node counts are reported separately for the left and right hemispheres. 
        The stimulated region (SP, superior parietal cortex, right hemisphere) is highlighted in bold, with the \num{27} nodes within this region stimulated simultaneously.
    }
    \label{tab:regions}
    \footnotesize
    \begin{tabular}{llcc}
        \toprule
        \multirow{2}{*}{\textbf{Label}}    &    \multirow{2}{*}{\textbf{Description}}    &    \multicolumn{2}{c}{\textbf{Number of nodes}}    \\
                                           &                                             &    \textbf{Left}    &    \textbf{Right}            \\
        \midrule
        \rowcolor{gray!15}
        BSTS                               &    Bank of the superior temporal sulcus     &    5                &    7                         \\
        CAC                                &    Caudal anterior cingulate cortex         &    4                &    4                         \\
        \rowcolor{gray!15}
        CMF                                &    Caudal middle frontal cortex             &    13               &    13                        \\
        CUN                                &    Cuneus                                   &    8                &    10                        \\
        \rowcolor{gray!15}
        ENT                                &    Entorhinal cortex                        &    3                &    2                         \\
        FP                                 &    Frontal pole                             &    2                &    2                         \\
        \rowcolor{gray!15}
        FUS                                &    Fusiform gyrus                           &    22               &    22                        \\
        IP                                 &    Inferior parietal cortex                 &    25               &    28                        \\
        \rowcolor{gray!15}
        IT                                 &    Inferior temporal cortex                 &    17               &    19                        \\
        ISTC                               &    Isthmus of the cingulate cortex          &    8                &    8                         \\
        \rowcolor{gray!15}
        LOCC                               &    Lateral occipital cortex                 &    22               &    19                        \\
        LOF                                &    Lateral orbitofrontal cortex             &    20               &    19                        \\
        \rowcolor{gray!15}
        LING                               &    Lingual gyrus                            &    16               &    17                        \\
        MOF                                &    Medial orbitofrontal cortex              &    12               &    12                        \\
        \rowcolor{gray!15}
        MT                                 &    Middle temporal cortex                   &    19               &    20                        \\
        PARC                               &    Paracentral lobule                       &    11               &    12                        \\
        \rowcolor{gray!15}
        PARH                               &    Parahippocampal cortex                   &    6                &    6                         \\
        PCAL                               &    Pericalcarine cortex                     &    9                &    10                        \\
        \rowcolor{gray!15}
        POPE                               &    Pars opercularis                         &    11               &    10                        \\
        PORB                               &    Pars orbitalis                           &    6                &    6                         \\
        \rowcolor{gray!15}
        PTRI                               &    Pars triangularis                        &    7                &    8                         \\
        PSTC                               &    Postcentral gyrus                        &    30               &    31                        \\
        \rowcolor{gray!15}
        PC                                 &    Posterior cingulate cortex               &    7                &    7                         \\
        PREC                               &    Precentral gyrus                         &    36               &    36                        \\
        \rowcolor{gray!15}
        PCUN                               &    Precuneus                                &    23               &    23                        \\
        RAC                                &    Rostral anterior cingulate cortex        &    4                &    4                         \\
        \rowcolor{gray!15}
        RMF                                &    Rostral middle frontal cortex            &    19               &    22                        \\
        SF                                 &    Superior frontal cortex                  &    50               &    46                        \\
        \rowcolor{gray!15}
        \textbf{SP}                        &    \textbf{Superior parietal cortex}        &    \textbf{27}      &    \textbf{27}               \\
        ST                                 &    Superior temporal cortex                 &    29               &    28                        \\
        \rowcolor{gray!15}
        SMAR                               &    Supramarginal gyrus                      &    19               &    16                        \\
        TP                                 &    Temporal pole                            &    4                &    3                         \\
        \rowcolor{gray!15}
        TT                                 &    Transverse temporal cortex               &    4                &    3                         \\
        \bottomrule
    \end{tabular}
\end{table}
\endgroup

\clearpage

\begingroup
\setlength{\tabcolsep}{7pt} 
\renewcommand{\arraystretch}{1.15} 
\begin{table}[p]
    \centering
    \caption{
        \textbf{TVB and Larter \& Breakspear parameters}.\\
        TUN column reports the parameters of the LB model tuned as described in the paper, starting from the default (DEF) configuration in~\citet{alstott2009modeling}. Parameter values that differ between the two configurations are indicated in bold, and reported before the unchanged ones. The first block of rows refers to the global parameters of the TVB simulator (notice the rescaling of some of them, as a straight consequence of the time rescaling in single-node equations), followed by parameters specific to the LB model.
    }
    \label{tab:params}
    \footnotesize
    \begin{tabular}{llcc}
        \toprule
        \multirow{2}{*}{\textbf{Parameter}}    &    \multirow{2}{*}{\textbf{Description}}                    &    \multicolumn{2}{c}{\textbf{Values}}                                                                        \\
                                               &                                                             &    \textbf{TUN}                                       &    \textbf{DEF}                                       \\
        \midrule
        \rowcolor{gray!15}
        $\boldsymbol{G}$                       &    \textbf{Global coupling strength}                        &    \textbf{3}                                         &    \textbf{1}                                         \\
        $\boldsymbol{c_v}$                     &    \textbf{Conduction velocity}                             &    $\boldsymbol{3 \cdot t_{\mathrm{scale}}}$          &    $\boldsymbol{3 \cdot t_{\mathrm{scale}}}$          \\
        \rowcolor{gray!15}
        $\boldsymbol{\sigma_{\xi}}$                     &    \textbf{Heun integrator noise}                           &    $\boldsymbol{10^{-7} \cdot t_{\mathrm{scale}}}$    &    $\boldsymbol{10^{-7} \cdot t_{\mathrm{scale}}}$    \\
        \midrule
        $\boldsymbol{t_{\mathrm{scale}}}$      &    \textbf{Timescale factor}                               &    \textbf{0.6}                                       &    \textbf{1}                                         \\
        \rowcolor{gray!15}
        $\boldsymbol{V_{\mathrm{Ca}}}$         &    \textbf{Ca Nernst potential}                             &    \textbf{1.1}                                       &    \textbf{1.0}                                       \\
        $\boldsymbol{a_{e \to e}}$             &    \textbf{Excitatory-to-excitatory synaptic strength}      &    \textbf{0.5}                                       &    \textbf{0.4}                                       \\
        \rowcolor{gray!15}
        $\boldsymbol{g_{\mathrm{K}}}$          &    \textbf{Conductance of K channels population}            &    \textbf{2.5}                                       &    \textbf{2.0}                                       \\
        $\boldsymbol{g_{\mathrm{L}}}$          &    \textbf{Conductance of leak channels population}         &    \textbf{1.1}                                       &    \textbf{0.5}                                       \\
        \rowcolor{gray!15}
        $\boldsymbol{a_{n \to i}}$             &    \textbf{Non-specific-to-inhibitory synaptic strength}    &    \textbf{L: 0.4023; R: 0.4}                         &    \textbf{0.4}                                       \\
        $\boldsymbol{T_{\mathrm{Na}}}$         &    \textbf{Threshold potential for Na channels}             &    \textbf{0.26}                                      &    \textbf{0.30}                                      \\
        \rowcolor{gray!15}
        $\boldsymbol{\delta_{\mathrm{K}}}$     &    \textbf{Variance of K channels threshold}                &    \textbf{0.4}                                       &    \textbf{0.3}                                       \\
        $\boldsymbol{V_{\mathrm{T}}}$          &    \textbf{Threshold potential for excitatory neurons}      &    \textbf{-0.1}                                      &    \textbf{0.0}                                       \\
        \rowcolor{gray!15}
        $\boldsymbol{\delta_{\mathrm{Z}}}$     &    \textbf{Variance of inhibitory threshold}                &    \textbf{0.60}                                      &    \textbf{0.65}                                      \\
        $g_{\mathrm{Ca}}$                      &    Conductance of population of Ca channels                 &    1.1                                                &    1.1                                                \\
        \rowcolor{gray!15}
        $r_{\mathrm{NMDA}}$                    &    Ratio of NMDA to AMPA receptors                          &    0.25                                               &    0.25                                               \\
        $g_{\mathrm{Na}}$                      &    Conductance of population of Na channels                 &    6.7                                                &    6.7                                                \\
        \rowcolor{gray!15}
        $V_{\mathrm{Na}}$                      &    Na Nernst potential                                      &    0.53                                               &    0.53                                               \\
        $V_{\mathrm{K}}$                       &    K Nernst potential                                       &    -0.7                                               &    -0.7                                               \\
        \rowcolor{gray!15}
        $V_{\mathrm{L}}$                       &    Nernst potential of leak channels                        &    -0.5                                               &    -0.5                                               \\
        $a_{i \to e}$                          &    Inhibitory-to-excitatory synaptic strength               &    2.0                                                &    2.0                                                \\
        \rowcolor{gray!15}
        $a_{n \to e}$                          &    Non-specific-to-excitatory synaptic strength             &    1.0                                                &    1.0                                                \\
        $I$                                    &    Subcortical input strength                               &    0.3                                                &    0.3                                                \\
        \rowcolor{gray!15}
        $b$                                    &    Time constant scaling factor                             &    0.1                                                &    0.1                                                \\
        $a_{e \to i}$                          &    Excitatory-to-inhibitory synaptic strength               &    1.0                                                &    1.0                                                \\
        \rowcolor{gray!15}
        $T_{\mathrm{Ca}}$                      &    Threshold potential for Ca channels                      &    -0.01                                              &    -0.01                                              \\
        $T_{\mathrm{K}}$                       &    Threshold potential for K channels                       &    0                                                  &    0                                                  \\
        \rowcolor{gray!15}
        $\delta_{\mathrm{Ca}}$                 &    Variance of Ca channels threshold                        &    0.15                                               &    0.15                                               \\
        $\delta_{\mathrm{Na}}$                 &    Variance of Na channels threshold                        &    0.15                                               &    0.15                                               \\
        \rowcolor{gray!15}
        $\phi$                                 &    Temperature scaling factor                               &    0.7                                                &    0.7                                                \\
        $\tau_{\mathrm{K}}$                    &    Time constant for K relaxation time                      &    1                                                  &    1                                                  \\
        \rowcolor{gray!15}
        $Q_{\mathrm{V}_{\mathrm{max}}}$        &    Max firing rate for excitatory populations               &    1                                                  &    1                                                  \\
        $\delta_{\mathrm{V}}$                  &    Variance of excitatory threshold                         &    0.65                                               &    0.65                                               \\
        \rowcolor{gray!15}
        $Q_{\mathrm{Z}_{\mathrm{max}}}$        &    Max firing rate inhibitory populations                   &    1                                                  &    1                                                  \\
        $Z_{\mathrm{T}}$                       &    Threshold potential for inhibitory neurons               &    0                                                  &    0                                                  \\
        \rowcolor{gray!15}
                                               &    Strength of excitatory coupling and balance              &                                                       &                                                       \\
        \rowcolor{gray!15}
        \multirow{-2}{*}{$C$}                  &    between internal and global dynamics                     &    \multirow{-2}{*}{0.1}                              &    \multirow{-2}{*}{0.1}                              \\
        \bottomrule
    \end{tabular}
\end{table}
\endgroup

\end{document}